\documentclass{emulateapj}

\usepackage{amssymb,amsmath}
\usepackage{amsfonts}
\usepackage{natbib}
\usepackage{longtable, graphicx, subfigure, epstopdf, multirow, setspace, verbatim, lmodern} 
\usepackage{aas_macros}
\newcounter{address}
\newcommand{\Msun}{\ifmmode {M_{\odot}}\else${M_{\odot}}$\fi}
\newcommand{\Rsun}{\ifmmode {R_{\odot}}\else${R_{\odot}}$\fi}
\newcommand{\lapprox }{{\lower0.8ex\hbox{$\buildrel <\over\sim$}}}
\newcommand{\gapprox }{{\lower0.8ex\hbox{$\buildrel >\over\sim$}}}
\def\amin{\ifmmode^{\prime}\else$^{\prime}$\fi}
\def\asec{\ifmmode^{\prime\prime}\else$^{\prime\prime}$\fi}

\shorttitle{Microlensing \& PTF}
\shortauthors{Price-Whelan et al.}
\begin{document}

\title{Statistical Searches for Microlensing Events in Large, Non-Uniformly Sampled Time-Domain Surveys: A Test Using Palomar Transient Factory Data}
\author{Adrian~M.~Price-Whelan\altaffilmark{1} Marcel~A.~Ag\"ueros\altaffilmark{1}, Amanda P. Fournier\altaffilmark{2,3}, Rachel Street\altaffilmark{3}, Eran O. Ofek\altaffilmark{4}, Kevin~R.~Covey\altaffilmark{5}, David Levitan\altaffilmark{6}, Russ R.~Laher\altaffilmark{7}, Branimir Sesar\altaffilmark{6}, Jason Surace\altaffilmark{7}}

\altaffiltext{1}{Department of Astronomy, Columbia University, 550 W 120th St., New York, NY 10027, USA; adrn@astro.columbia.edu}
\altaffiltext{2}{Department of Physics, Broida Hall, University of California, Santa Barbara, CA 93106, USA}
\altaffiltext{3}{Las Cumbres Observatory Global Telescope Network, Inc., 6740 Cortona Dr.\ Suite 102, Santa Barbara, CA 93117, USA}
\altaffiltext{4}{Benoziyo Center for Astrophysics, Weizmann Institute of Science, 76100 Rehovot, Israel}
\altaffiltext{5}{Lowell Observatory, 1400 West Mars Hill Rd, Flagstaff, AZ 86001, USA}
\altaffiltext{6}{Division of Physics, Mathematics, and Astronomy, California Institute of Technology, Pasadena, CA 91125, USA}
\altaffiltext{7}{Spitzer Science Center, California Institute of Technology, Mail Stop 314-6, Pasadena, CA 91125, USA}

\begin{abstract}
Many photometric time-domain surveys are driven by specific goals, such as searches for supernovae or transiting exoplanets, which set the cadence with which fields are re-imaged. In the case of the Palomar Transient Factory (PTF), several sub-surveys are conducted in parallel, leading to non-uniform sampling over its $\sim$$20,000~\mathrm{deg}^2$ footprint. While the median $7.26~\mathrm{deg}^2$ PTF field has been imaged $\sim$40 times in \textit{R}-band, $\sim$$2300~\mathrm{deg}^2$ have been observed $>$100 times. We use PTF data to study the trade-off between searching for microlensing events in a survey whose footprint is much larger than that of typical microlensing searches, but with far-from-optimal time sampling. To examine the probability that microlensing events can be recovered in these data, we test statistics used on uniformly sampled data to identify variables and transients. We find that the von Neumann ratio performs best for identifying simulated microlensing events in our data. We develop a selection method using this statistic and apply it to data from fields with $>$10 $R$-band observations, $1.1\times10^9$ light curves, uncovering three candidate microlensing events. We lack simultaneous, multi-color photometry to confirm these as microlensing events. However, their number is consistent with predictions for the event rate in the PTF footprint over the survey's three years of operations, as estimated from near-field microlensing models. This work can help constrain all-sky event rate predictions and tests microlensing signal recovery in large data sets, which will be useful to future time-domain surveys, such as that planned with the Large Synoptic Survey Telescope.
\end{abstract}


\section{Introduction}
Over the past 20 years, microlensing, in which gravitational lensing causes a transient increase in the flux from a background point source, has been used to search for dark and compact objects \citep{original_macho, oslowski2008, sartore2010}, to study Galactic structure and kinematics \citep{binney2000}, to determine the shape of stars \citep{rattenbury2005}, and to identify extrasolar planets \citep[][and references therein]{gaudi2011}. Such studies were once limited by the small number of detected events, but thanks to advances in CCD technology and the development of dedicated microlensing surveys, a few thousand events are now observed each year.\footnote{e.g., \url{http://ogle.astrouw.edu.pl/ogle4/ews/ews.html}.} In the coming decade, this number is expected to increase as the next generation of photometric, time-domain surveys comes online, providing an opportunity for the precise study of otherwise hard-to-characterize objects, such as low-mass stars, sub-stellar objects, and isolated neutron stars. 

Microlensing surveys typically focus on high-density stellar regions, such as the Galactic Bulge, M31, or the Magellanic Clouds \citep[e.g.,][]{original_ogle, original_macho, eros_original, crotts1996}, but microlensing events are not limited to dense stellar fields. Indeed, lensing events away from these fields are potentially very interesting. While the microlensing event rate in low-density stellar regions will be smaller, compared to events found by modern microlensing surveys, these are expected to involve closer sources and lenses, and less crowded backgrounds \citep{mesolensing1}. Thus, such lenses offer more opportunities to constrain the properties of the events.

\begin{figure*}[!t]
\centering\includegraphics[scale=0.4,trim=0 0 0 27, clip]{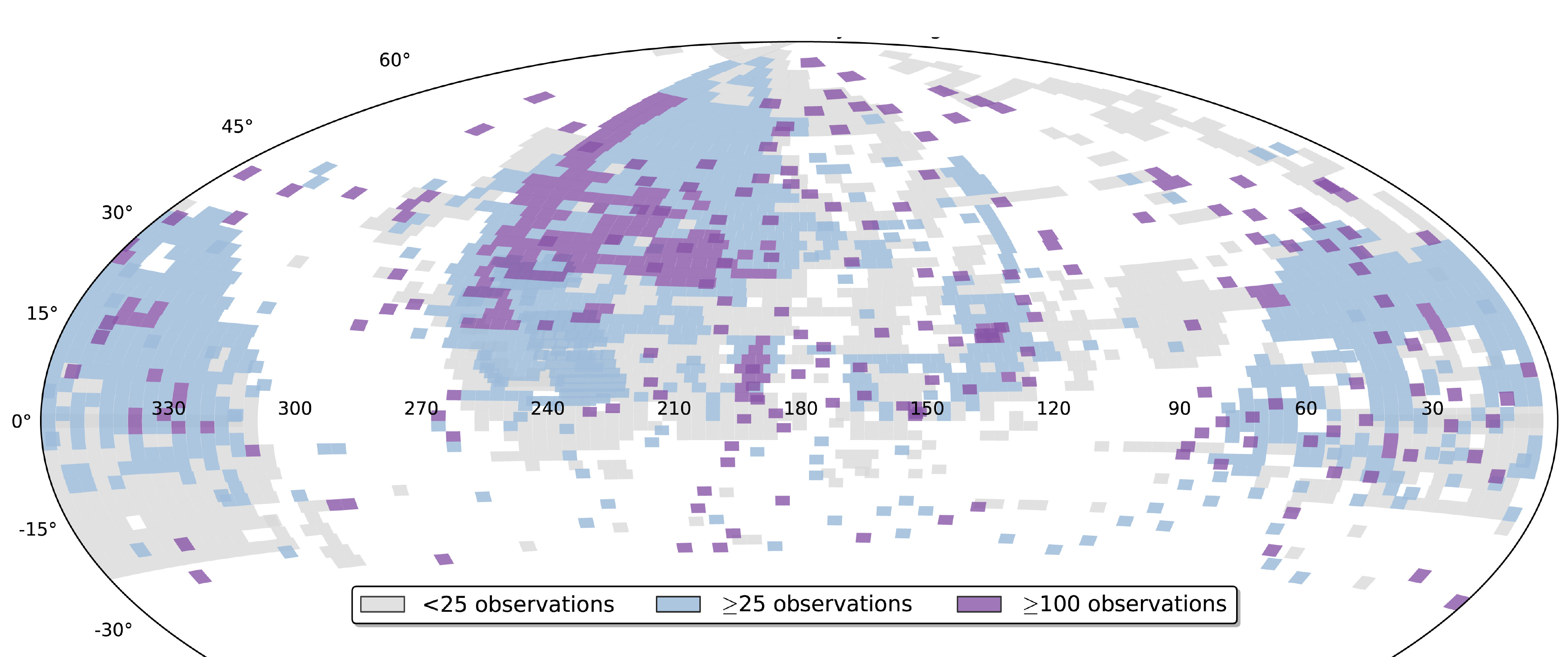}
\caption{PTF $R$-band survey footprint, in equatorial coordinates. The fields are color-coded by the total number of observations. The field size corresponds to the actual area covered by a single PTF exposure.}\label{fig:survey_footprint}
\end{figure*}	

Only one event outside of a dense stellar field has been recorded to date: the Tago event \citep{fukui2007, gaudi2008}, a serendipitous detection of the microlensing of a bright source. \cite{gaudi2008} set probabilistic limits on the mass, distance, proper motion, and magnitude of this lens using existing observations near the event's location, and further observations may turn these limits into precise measurements. Such an analysis is usually not possible for events in high-density fields where the lenses are, on average, much farther away and blending makes photometry --- and, therefore, mass measurements --- challenging. By contrast, high-Galactic-latitude events probe lenses with distances $\lesssim$1~kpc; the Einstein radii, therefore, tend to be larger ($\sim$milliarcseconds) and may cause detectable astrometric signatures \citep[][]{han2008, mesolensing2, mesolensing1}. A measurement of the astrometric shift due to microlensing will break the distance-mass degeneracy in the lensing parameters and enable a direct measurement of the lens mass. Finding even a handful of nearby lenses that can be studied in the same kind of detail as the Tago event would, therefore, be very valuable. 

In this paper, we use the Palomar Transient Factory (PTF) survey, described in Section~\ref{sec:ptf}, as a test case for studying the frequency and detectability of microlensing events in all-sky, irregularly sampled, time-domain surveys. Although to date no microlensing event has been reported using PTF data, a simple extrapolation from previous all-sky rate estimates suggests $\sim$1--10 detectable microlensing events within the PTF footprint over the survey's three years of operation \citep{gaudi2008, han2008}. We examine the selection methods used by microlensing surveys to identify event signatures in light curves as well as an event identification procedure based on a set of variability statistics. We then present the detection efficiency for these methods as computed for a representative set of PTF fields (Section~\ref{sec:event_recovery}). In Section~\ref{sec:search}, we apply a set of criteria that use the results of these tests to all PTF light curves with $>$10 high-quality observations to identify candidate microlensing events. We discuss eight interesting transients and three candidate events identified in this manner before concluding in Section~\ref{sec:conc}. 

\section{PTF: Survey Design and Consequences for Microlensing Searches}\label{sec:ptf} 
PTF data are collected using the former Canada-France-Hawaii Telescope 12K$\times$8K mosaic camera, which has 11 working chips, $0.92\times10^8$ pixels, and a 7.26 deg$^2$ field-of-view \citep{rahmer2008}, mounted on the 48-inch Oschin Schmidt Telescope (the P48) at Palomar Observatory, CA. Under median seeing conditions (2.2$\arcsec$), observations in Mould $R$ or Sloan Digital Sky Survey \citep[SDSS;][]{york00} $g$ achieve 2.0$\arcsec$ full-width half-maximum images and reach 5$\sigma$ magnitudes of $R \approx 21.0$ and $g \approx 21.3$ mag in each 60~s exposure \citep{nick2009,rau2009,nick2010}. 

While a real-time image-differencing pipeline identifies transients of interest and passes these to a dedicated photometric follow-up telescope, a separate pipeline generates PTF light curves. Images are processed using standard reduction procedures including de-biasing, flat-fielding, and astrometric calibration (Laher et al., in prep). SExtractor is used for source identification \citep{bertin96}. The absolute photometric calibration of the images is good to a systematic limit of $\sim$2\% for photometric nights, as described in \cite{ofek2012}.  The final light curves are produced using relative photometric calibration, which refines the calibration to $<$1\% for photometric nights and improves the calibration for bad nights \citep[Levitan et al., in prep; for algorithm details see][]{levitan2011, ofek2012}. The resulting archive is an excellent resource for the study of periodic forms of variability \citep[e.g., stellar or asteroid rotation;][]{agueros11,polishook2012}. 

\begin{figure}[h]
\begin{center}
\includegraphics[width=0.47\textwidth, trim=2 2 5 5, clip]{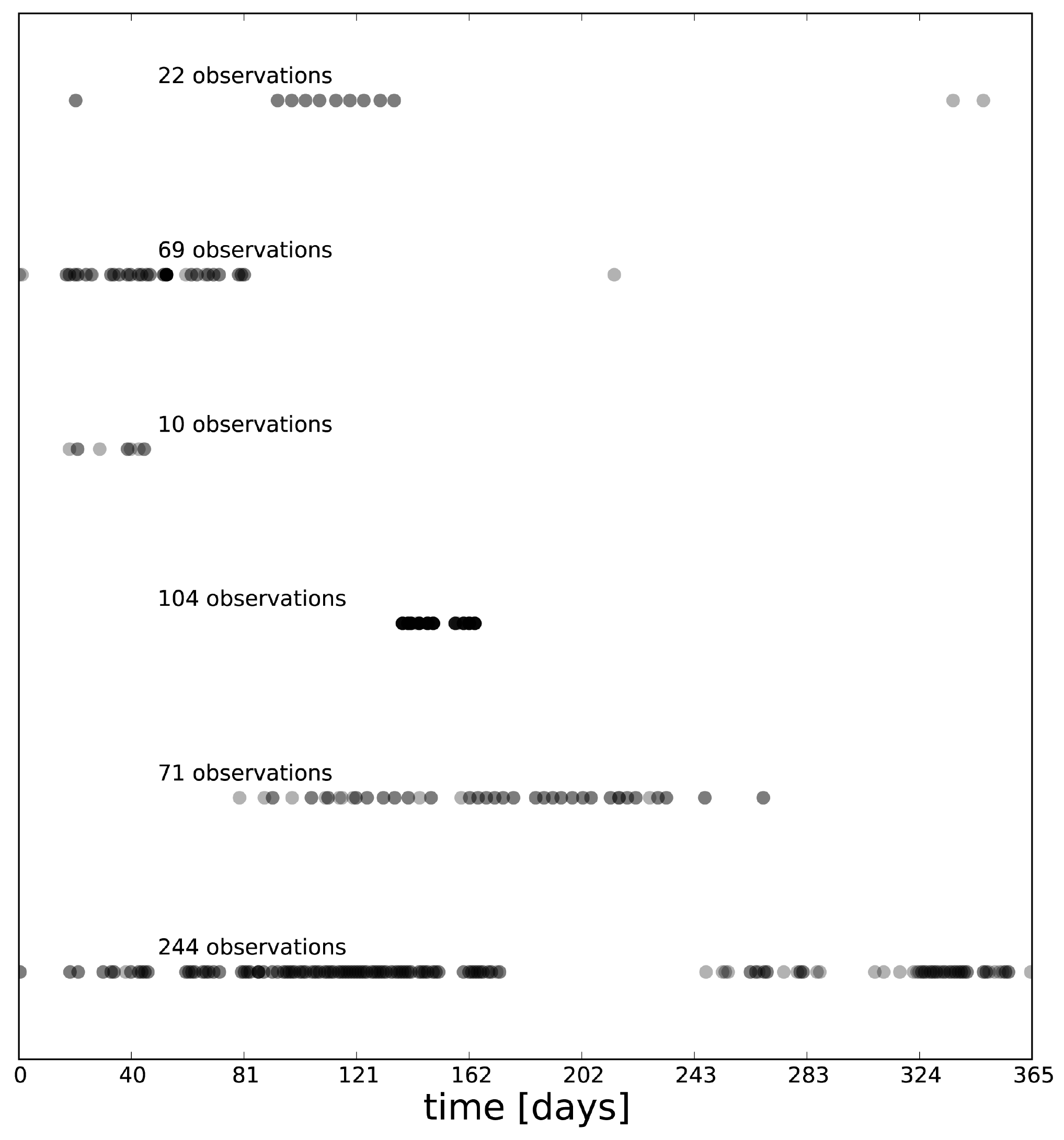}
\caption{Sampling of six different PTF fields over the course of a year. Darker points indicate that more exposures were taken that night; the total number of observations for each field is indicated.}\label{fig:sampling}
\end{center}
\end{figure}

As of 2012 Dec, the PTF footprint includes $\sim$15,224 (2766) deg$^2$ imaged $>$10 ($>$100) times in $R$-band and $\sim$5430 (290) deg$^2$ imaged that often in $g$-band (see Figure~\ref{fig:survey_footprint}). The PTF survey footprint is not uniformly sampled either spatially or temporally. Each field has a unique sampling pattern determined by which of the PTF sub-surveys it belongs to, what time of the year it is visible, and how high a priority it is given by the scheduler. The imaging cadences range from 1-5 days, but other cadences are possible: for example, there was a higher-cadence campaign of a field in Orion to find transiting planets around young stars \citep{vaneyken2011}. The result is that PTF light curves often contain gaps and regions of high-cadence observations and/or of low-cadence observations, so that the archive is a massive dataset of irregularly-sampled, time-domain photometry. Figure~\ref{fig:sampling} shows six randomly selected light curves and illustrates the varying cadences and coverage that different fields may have over the same one-year period. While the PTF database is not yet public, we have made $\sim$$10^4$ randomly selected light curves available at \url{http://adrian.pw/ptf}. These light curves meet the quality criteria described in Section~\ref{sec:detection_eff} and can be used to test the statistical methods described here (or others!).

\begin{figure}[t]
\centering\includegraphics[width=0.5\textwidth, trim=21 20 25 70, clip]{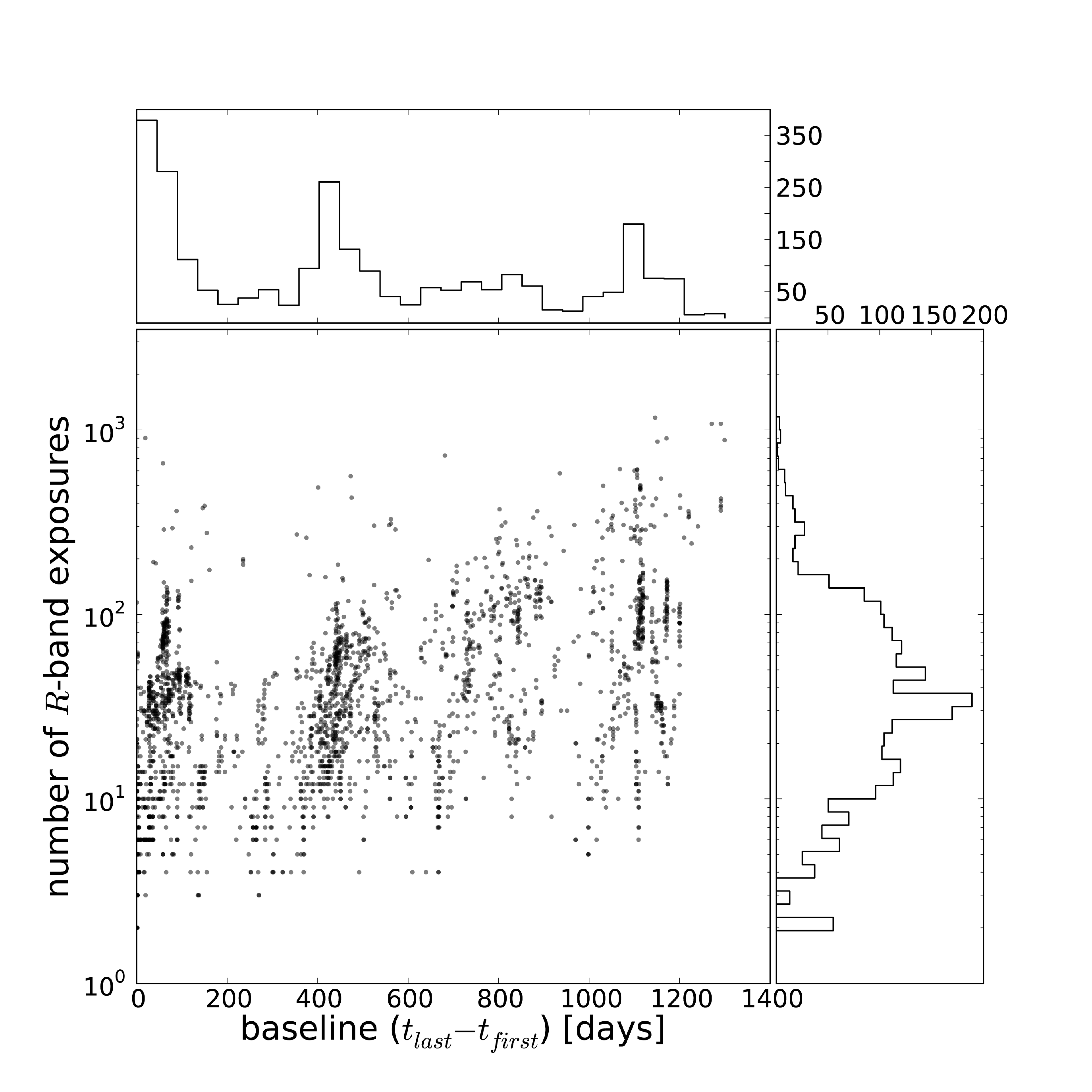}
\caption{Total number of $R$-band exposures and baselines (number of days between first and last exposure) for all PTF fields. The histogram to the right shows that the exposure distribution is peaked at 30-40; the histogram at the top shows that the baseline distribution has three peaks, corresponding to fields observed over $\lapprox$weeks, $\sim$1 year, and $\sim$3 years (the full length of the survey).}
\label{fig:num_obs}
\end{figure}

\begin{figure}[!h]
\centering\includegraphics[width=0.48\textwidth, trim=5 0 0 7, clip]{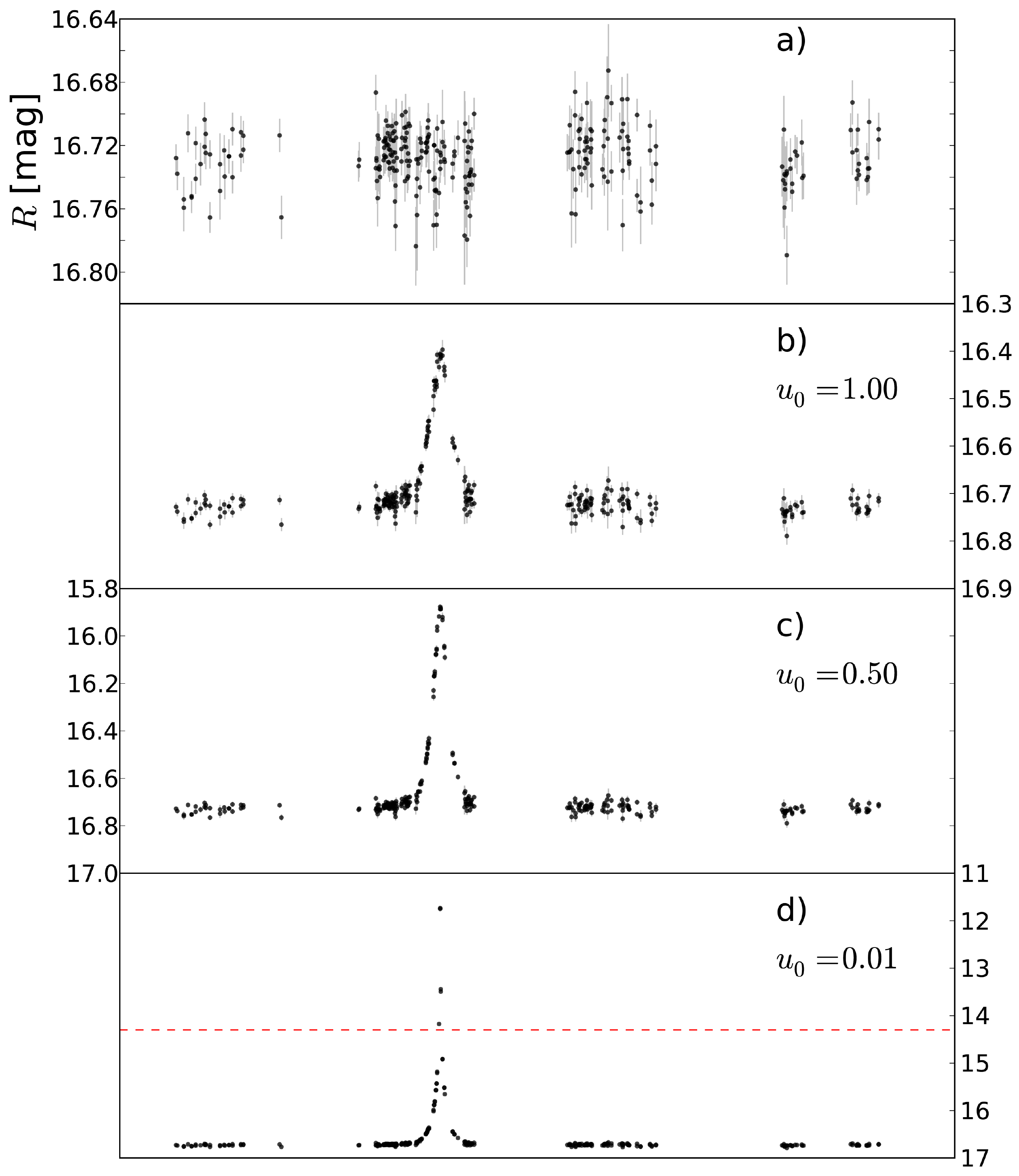}
\caption{Simulating the effect of $20$-day microlensing events with impact parameters $u_0 = 1.0$, 0.5, and 0.01, on a random PTF $R$-band light curve. The original light curve is shown at the top. The dashed line in the bottom panel shows the approximate saturation limit of the PTF camera in the $R$-band; such an event with a brighter source might therefore be missed by the survey.}\label{fig:microlensing_sim}
\end{figure}

In Figure~\ref{fig:num_obs}, we plot the number of $R$-band exposures for each field against the observational baseline (the number of days between the first and last exposure). Given the exposure distribution, which peaks at 30-40 but includes a long tail to larger numbers, we restrict our sample to the light curves with $>$10 \textit{R}-band observations.  

\section{Microlensing event recovery} \label{sec:event_recovery}
Microlensing surveys typically use difference image analysis \citep{alard1998} to identify transient events in raw imaging data. The light curves of transient sources are then analyzed and vetted using a variety of selection methods to search for microlensing event candidates and distinguish them from, e.g., variable stars, outbursting systems, and novae. Surveys have approached this process differently (as described in \citealt{ogle_optical_depth, con_idx, alcock2000, macho_detection_efficiency, udalski03,  hamadache2009, wyrzykowski2009, sumi2011}), but the general idea is to require that: 
\begin{enumerate}
	\item any selected light curve has some number of consecutive data points brighter than some threshold,
	\item compared to flat or linear light-curve models, a microlensing model best describe the data, and
	\item the microlensing model event parameters have physically reasonable values.
\end{enumerate}

What would a microlensing event look like in a typical PTF light curve? If the source is an unblended point source and the lens is a foreground, dim object, a microlensing event is fully described by three parameters: the angular impact parameter $u_0$, the peak time of the event $t_0$, and the timescale of the event (Einstein crossing time) $t_E$. In terms of a dimensionless projected distance between the source and lens (in units of Einstein radius), $u$, the amplification factor $A$ and flux $F$ as a function of time can be defined as: 
\begin{align}
	u(t) &= \sqrt{u_0^2 + 2\Big(\frac{t-t_0}{t_E}\Big)},\\
	A(t) &= \frac{u^2 + 2}{u\sqrt{u^2 + 4}},\\
	F(t) &= A(t)\times F_{source}.
\end{align}
\citep{paczynski1986}. The microlensing perturbation can also be expressed as:
\begin{align}\label{eq:ml_model}
	m(t) &= m_0(t) - 2.5\log A(t),
\end{align}
where $m_0(t)$ is the unperturbed, but possibly time-variable, magnitude of the source.

Even in cases of high amplification ($u_0<<1$), a survey may miss or poorly sample an event if $t_E$ is short ($t_E \lesssim\mathrm{days}$), while if $t_E$ is long ($t_E \gtrsim 1~\mathrm{year}$) the event may be confused with other forms of long-duration variability. Figure~\ref{fig:microlensing_sim} illustrates the effect of simulated microlensing events with a fixed $t_E=20~\mathrm{days}$, but different angular impact parameter, $u_0$, on a random well-sampled PTF light curve. 

Applying the standard microlensing-search prescription to the PTF data presents obvious challenges. For example, microlensing surveys have relatively uniform time sampling of their survey footprint over an observing season, justifying the first requirement. But it is harder to motivate such a cut on data with significant and irregular gaps. In order to develop the most successful procedure for PTF data, we therefore examine the relative performance of a set of variability indices and of the traditional prescription in selecting simulated microlensing events.

\begin{figure}[h]
\centering\includegraphics[width=.48\textwidth, trim=10 50 50 50, clip]{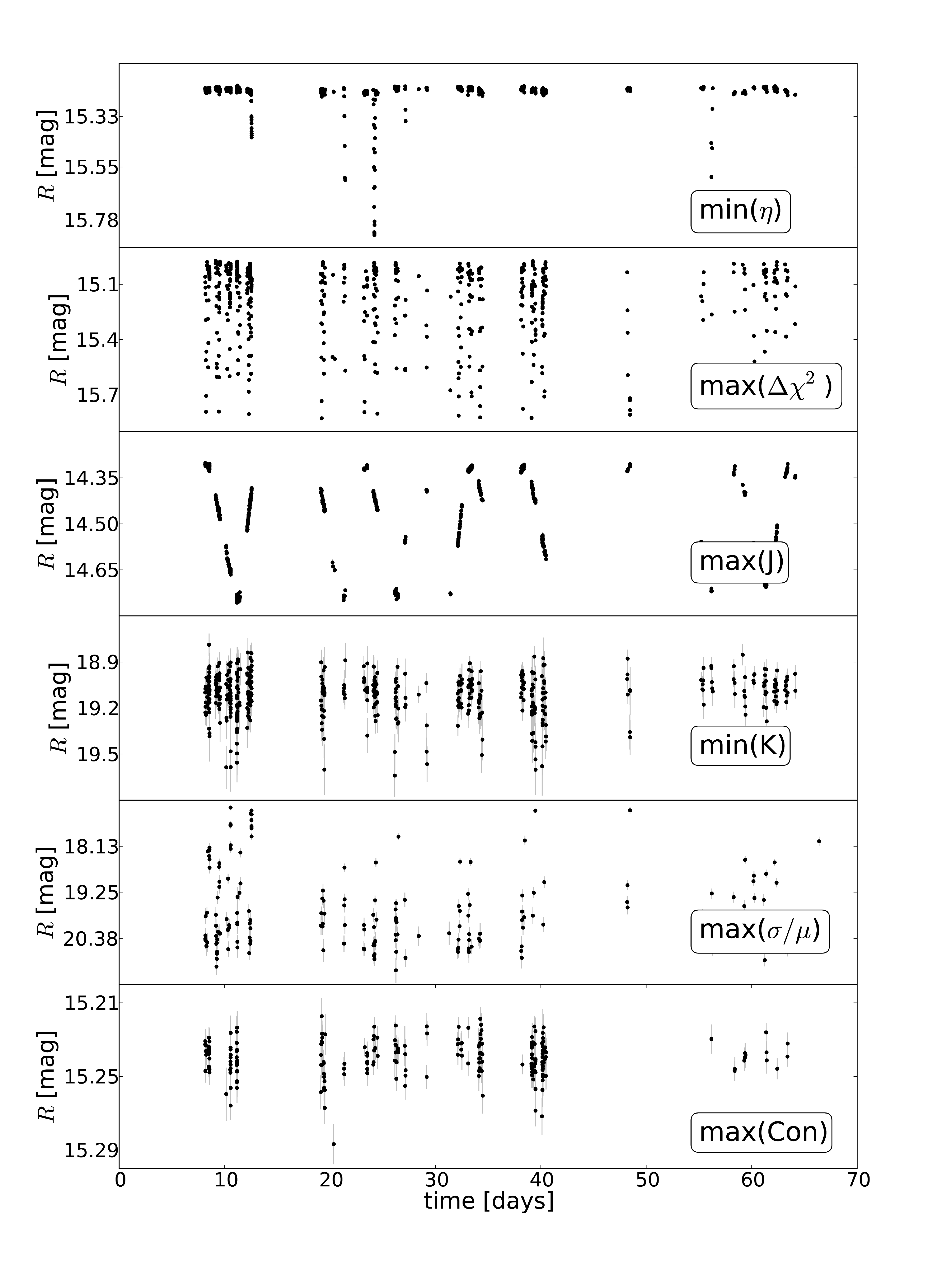}
\caption{Light curves selected from PTF field 3756 with maximally outlying values for each variability index. For $\eta$ and $K$, this corresponds to the light curve with the \emph{minimum} value of the index over the entire field. For the other indices, these are the light curves with the \emph{maximum} value of the relevant index over the field. These light curves illustrate the type of variability that each index is most sensitive to. } \label{fig:indices_examples}
\end{figure}

\subsection{The Variability Indices} 
We choose five statistical measures of variability compiled by \cite{shin2009} that have been previously applied to the classification and discovery of periodic variables. These variability indices are $\sigma/\mu$, $Con$, $\eta$, $J$, and $K$.\footnote{We do not implement the sixth index described by \cite{shin2009}, $AoVM$, because it mainly helps identify periodic sources.} $\sigma/\mu$ is the ratio of the sample standard deviation ($\sigma$) to the sample mean ($\mu$), 
\begin{align}
	\frac{\sigma}{\mu} = \frac{\sqrt{\sum^N_i (x_i - \mu)^2 / (N-1)}}{\sum^N_i x_i/N},
\end{align}
where N is the total number of observations. 

We modify the definition of $Con$ to be the number of clusters of three or more consecutive observations that are more than $3\sigma$ brighter than the reference magnitude of the source (e.g., for a single microlensing event in an otherwise flat light curve, $Con=1$). This change allows us to use the performance of $Con$ as a proxy for the consecutive-point requirement described above. 

$\eta$, the von Neumann ratio \citep[also known as the Durbin-Watson statistic;][]{von_neumann1941, durbin50}, is the mean square successive difference divided by the sample variance:
\begin{align}
	\eta = \frac{\delta^2}{\sigma^2} = \frac{\sum^{N-1}_i(x_{i+1} - x_i)^2/(N-1)}{\sigma^2}.
\end{align}
$\eta$ is small when there is strong positive serial correlation between successive data points. 

$J$ and $K$ were suggested by \cite{stetson1996}:
\begin{align}
	\delta_i &= \sqrt{\frac{N}{N-1}}\frac{x_i-\mu}{e_i},\\
	J &= \sum^{N-1}_i sign(\delta_i \delta_{i+1})\sqrt{|\delta_i \delta_{i+1}|},\\
	K &= \frac{1/N\sum^N_i |\delta_i|}{\sqrt{1/N\sum^N_i\delta_i^2}},
\end{align}
where $e_i$ is the photometric error of each data point, and the $sign$ function returns $\pm$1, depending on the sign of the argument. $J$ tends to 0 for non-variable stars, but is large when there are significant differences between successive data points in a light curve. $K$ is a measure of the kurtosis of the distribution of data points. We add one more index, $\Delta \chi^2$, the difference in $\chi^2$ between fitting a Gaussian model and fitting a linear model to a light curve. This is a standard statistical test used by microlensing surveys, and allows us to compare the relative performance of the (slightly modified) \cite{shin2009} indices and of this approach.

Specifically, we use a Levenberg-Marquardt optimizer to perform a least-squares fit with each of these models to the light curves and then compute $\Delta\chi^2=\chi_{linear} - \chi_{gaussian}$. Our tests with $\Delta \chi^2$ below compare the distribution of values over light curves on the same chip. As a result, the number of data points and number of model parameters are constant, and we therefore do not include an Akaike or Bayesian information criterion (AIC or BIC) term. For the Gaussian fit, the optimizer is initialized with a $\sigma$ of 10 days, centered on the brightest data point.

Figure~\ref{fig:indices_examples} shows maximally outlying light curves for each variability statistic selected from $\sim$20,000 light curves for objects on a single CCD in PTF field 3756. 
Clearly, the indices are sensitive to different aspects of variability in the data. We expect $\sigma/\mu$ to be most useful for discerning periodic or semi-periodic variability where the variance is large; the expectations are not as clear for other indices. 

\begin{figure}[h]
\centering
        \subfigure{
	   \includegraphics[width=.467\textwidth, trim=5 9 5 3, clip] {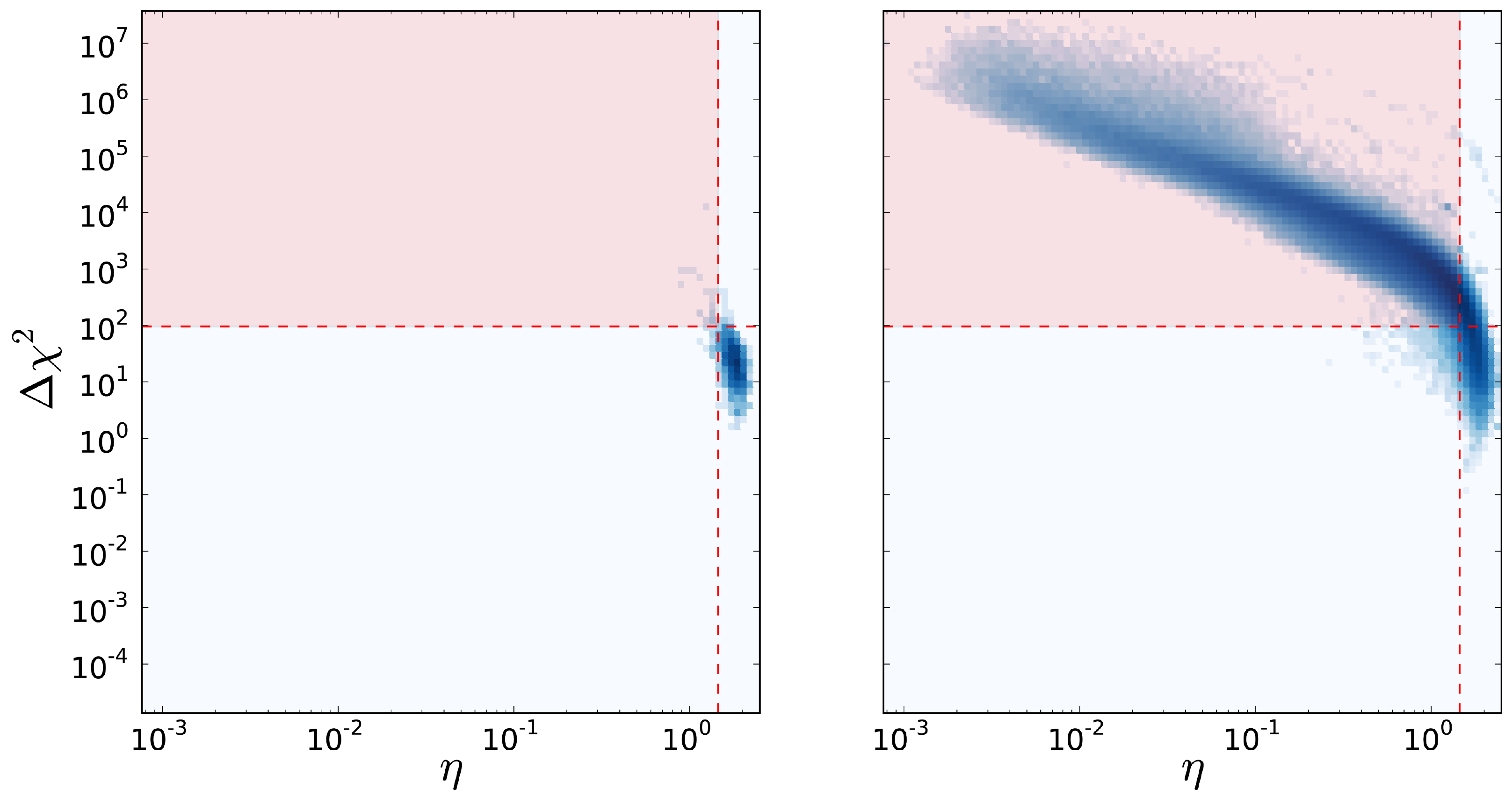}
	 }
	
	 \subfigure{
	   \includegraphics[width=.47\textwidth, trim=10 9 0 7, clip] {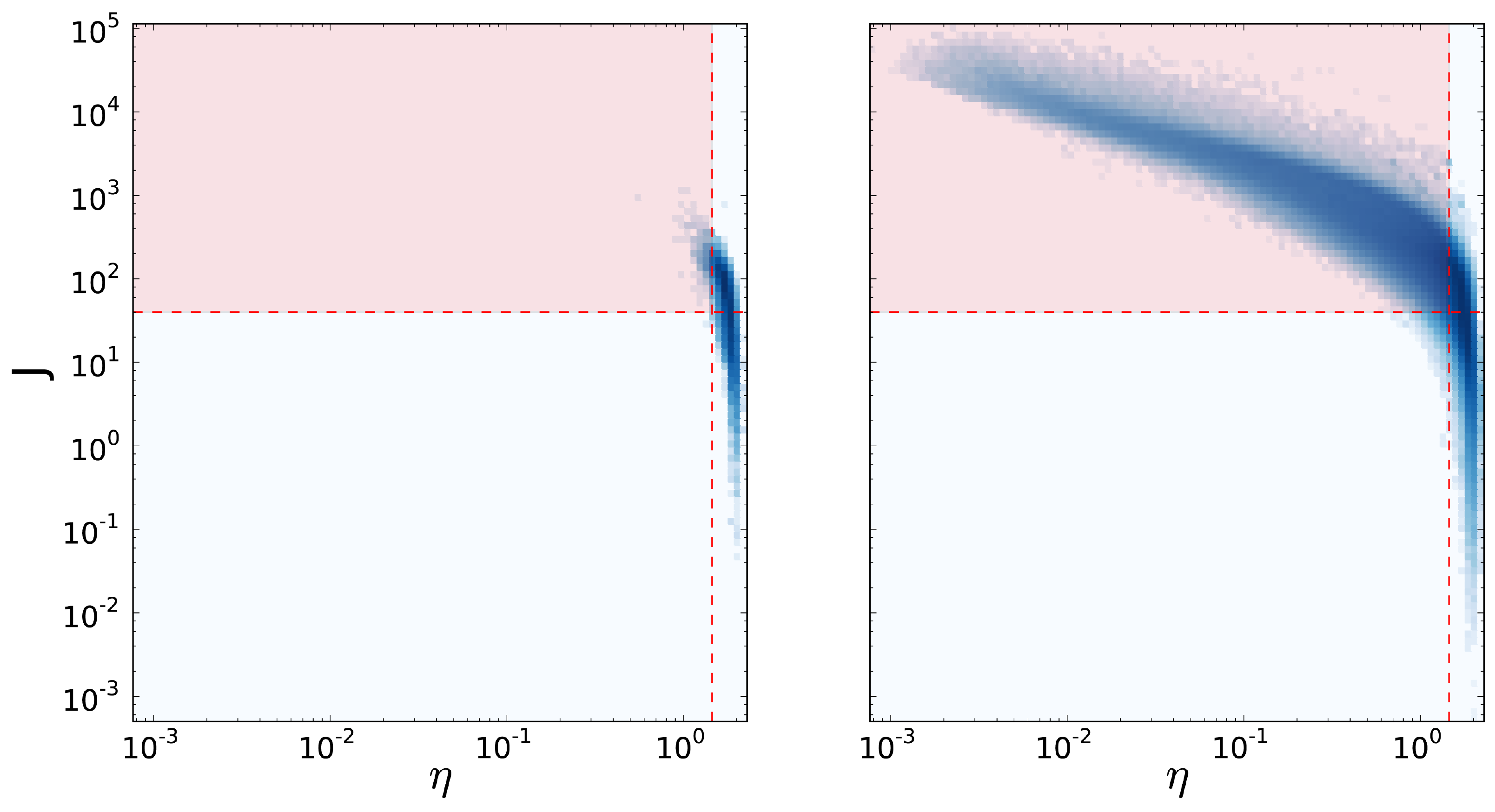}
	 }
	
	 \subfigure{
	   \includegraphics[width=.47\textwidth, trim=10 6 0 7, clip] {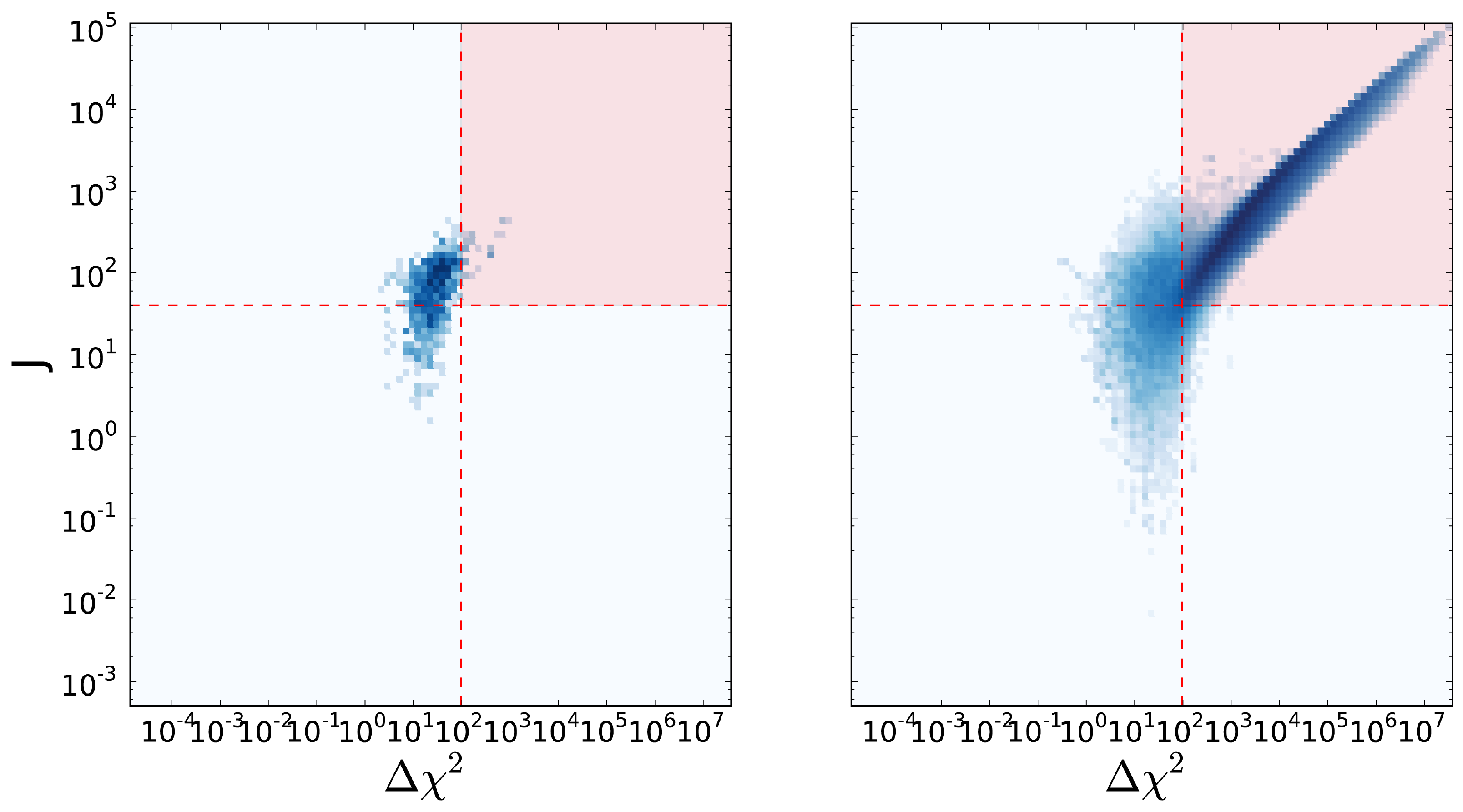}
	 }
	\caption{Two-dimensional density histograms for projections of the six-dimensional variability statistic distribution for 10,000 light curves from PTF field 100018 (left) and with the addition of simulated microlensing events (right). Red (dashed) lines are the 1\% false positive recovery selection boundaries for each index. Shaded (red) regions show the subspaces where microlensing events are expected to fall, as defined by the two statistics in question. } 
	\label{fig:var_indices}
\end{figure}

We conduct Monte Carlo simulations and inject artificial microlensing events into real PTF light curves. We then compare the distributions of variability indices for a set of light curves with and without these simulated events to determine the regions of parameter space where microlensed light curves fall. Figure~\ref{fig:var_indices} shows projections of the distribution for various combinations of the variability indices for another PTF field. These simulations do define regions to search for microlensing events, but there is still a large amount of overlap between the distributions with and without microlensing events. Below we describe our method for defining selection boundaries for each variability statistic to determine which is most efficient for detecting microlensing events.

\subsection{Event Selection and Detection Efficiency} \label{sec:detection_eff}
We begin by randomly sampling 1000 light curves from each chip in a given PTF field. At least half of the individual observations in these light curves must be defined as ``good'' \citep[see description of processing pipeline in][]{nick2009}. The rejected light curves are for objects that are either faint or near bright stars whose scattered light and diffraction spikes cause large photometric errors. 

For each of these light curves, we use the date and magnitude error information to simulate 100 light curves with purely Gaussian scatter. We find the value of each variability index such that the selection using that index to identify interesting light curves returns 1\% of these scrambled light curves --- i.e., we set the limiting value of each index such that the false positive recovery (FPR) rate is 1\% per trial. We again randomly sample 1000 light curves from each CCD with $>$10 good observations and compute the set of indices for these light curves. Finally, we add 100 different simulated microlensing events to each light curve and evaluate how often these events are recovered given the 1\% FPR cut defined above. 

\begin{figure}
\centering
	\subfigure{
	   \includegraphics[width=.45\textwidth, trim=0 8 6 3, clip] {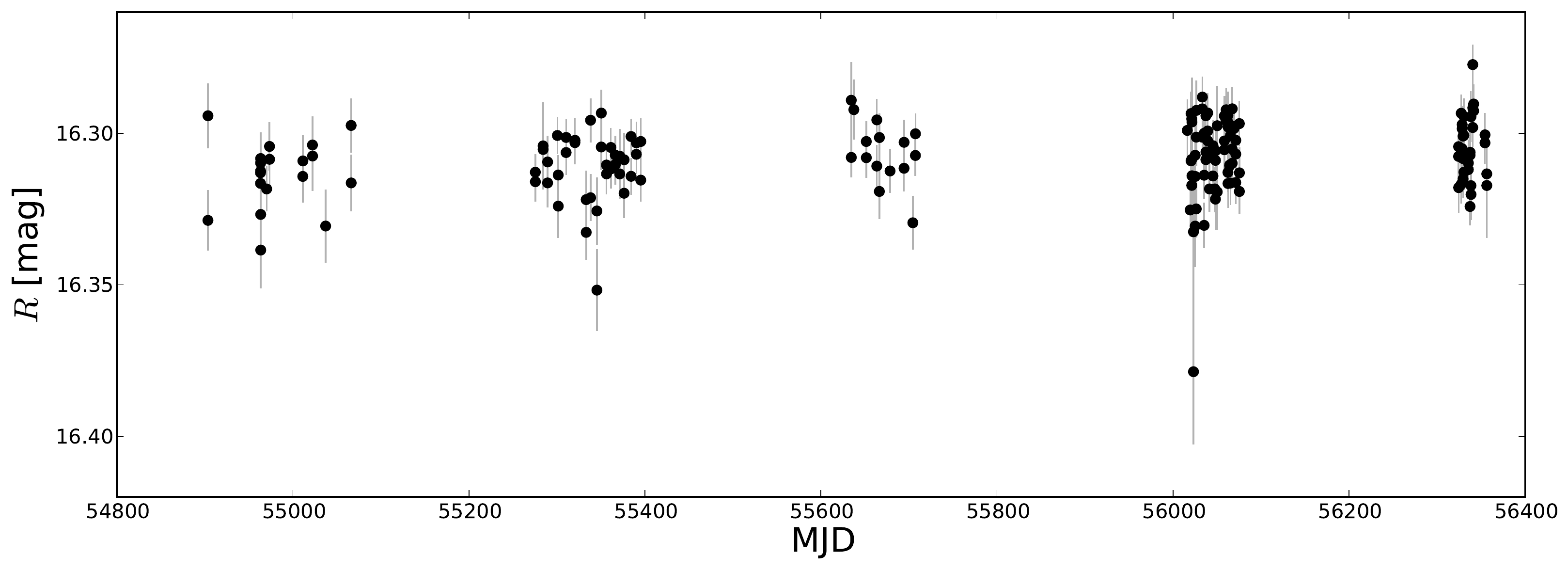}
	   \label{fig:subfig1}
	 }
	
	 \subfigure{
	   \includegraphics[width=.46\textwidth, trim=8 8 5 5, clip] {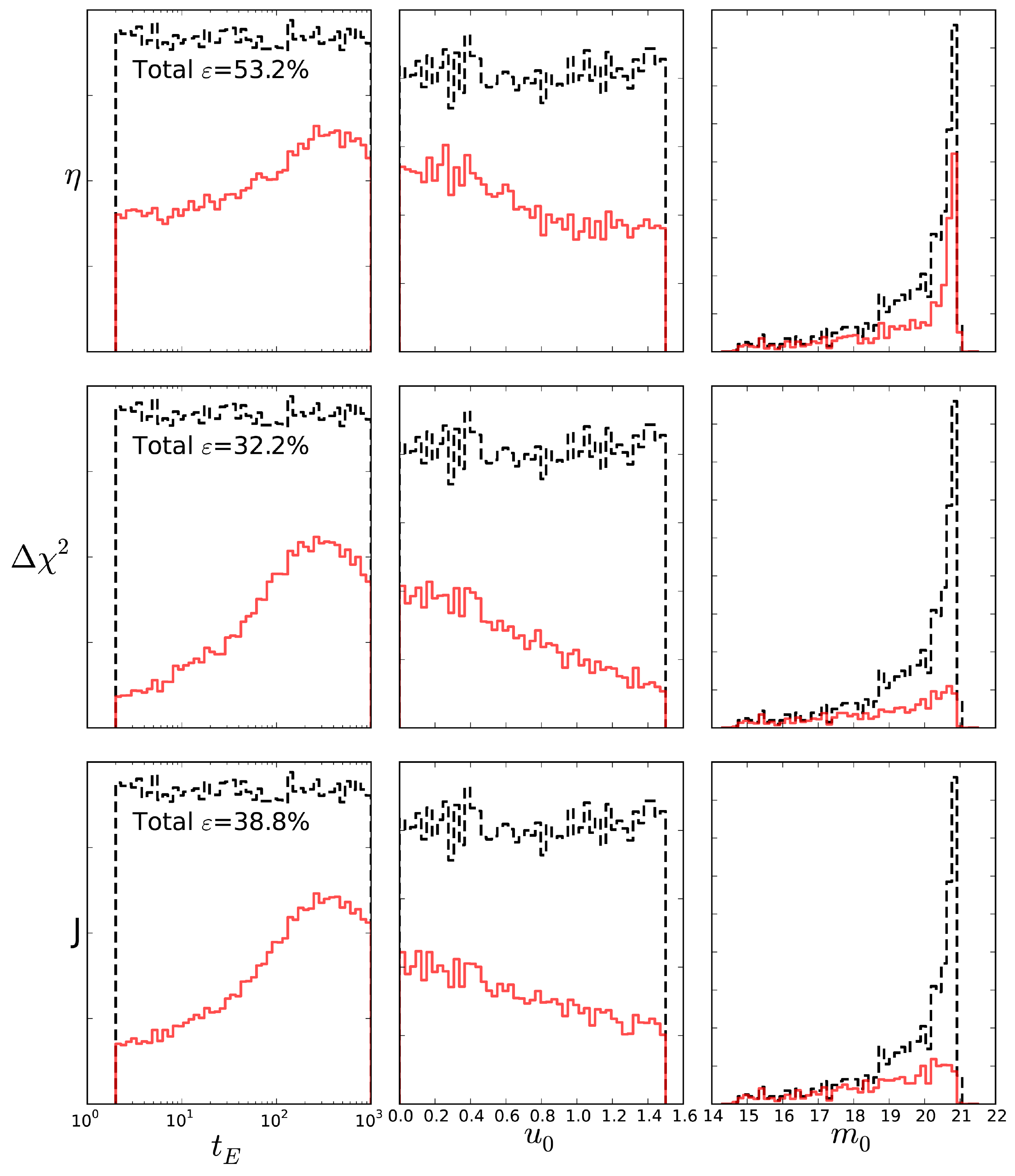}
	   \label{fig:subfig2}
	 }
\caption{{\it Top:} A randomly selected light curve from PTF field 4327. Note the sampling pattern. {\it Bottom:} Detection efficiency $\varepsilon$ for $\eta$, $\Delta\chi^2$, and $J$, as a function of the simulated microlensing event parameters $t_E$, $u_0$, and $m_0$ (the event timescale, impact parameter, and unmagnified source magnitude, respectively). Black (dashed) lines show the distributions for all light curves (normalized, so y-axis scale is arbitrary), red (solid) lines show the recovered distributions.}  \label{fig:detection_efficiency_4327}
\end{figure}

The parameters for each simulated event are chosen as follows: $t_0$ is drawn from a uniform distribution between the first and last observation date, $u_0$ is drawn from a uniform distribution between 0 and $\sim$1.5 \citep[the impact parameter that causes a maximum deviation larger than $\sim$$5\sigma$ for a $R\sim17.5$ mag source; e.g.,][]{macho_detection_efficiency}, and $t_E$ is drawn from a log-uniform distribution between 1 and 1000 days. For each iteration we recompute the variability indices and store the event parameters. 

\begin{figure}
\centering
        \subfigure{
	   \includegraphics[width=.45\textwidth, trim=0 8 6 3, clip] {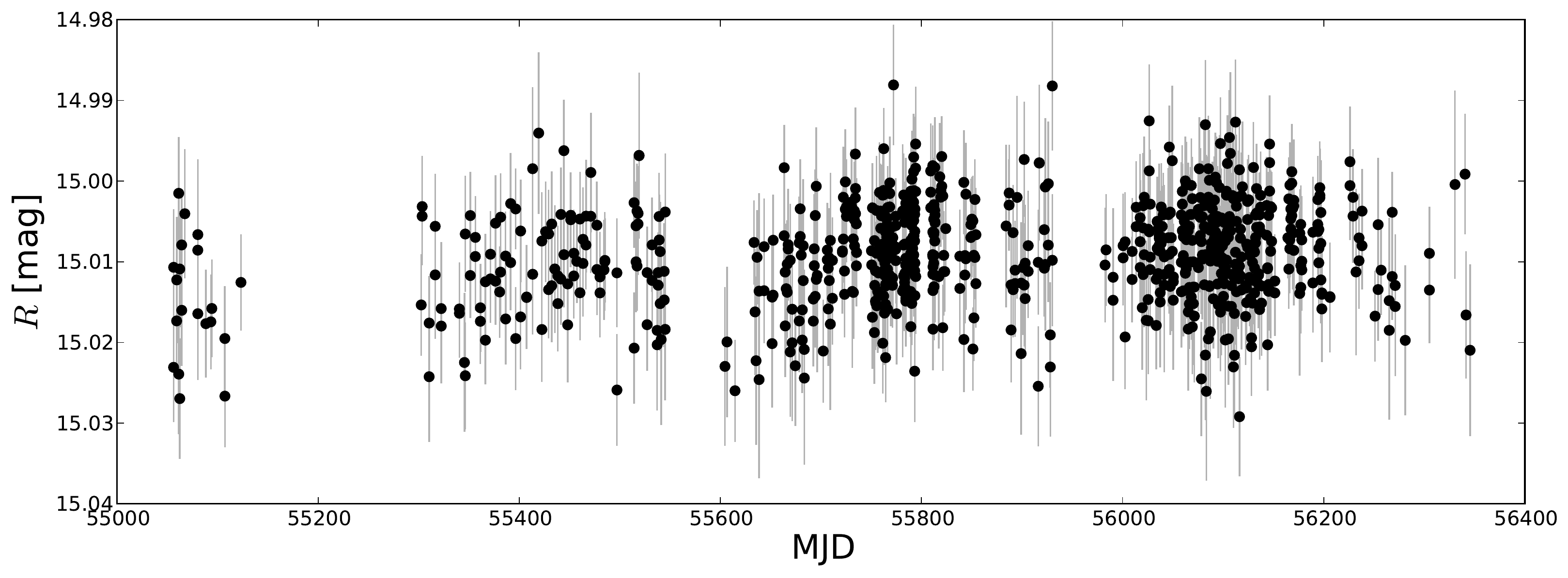}
	   \label{fig:subfig1}
	 }
	
	 \subfigure{
	   \includegraphics[width=.46\textwidth, trim=8 8 5 5, clip] {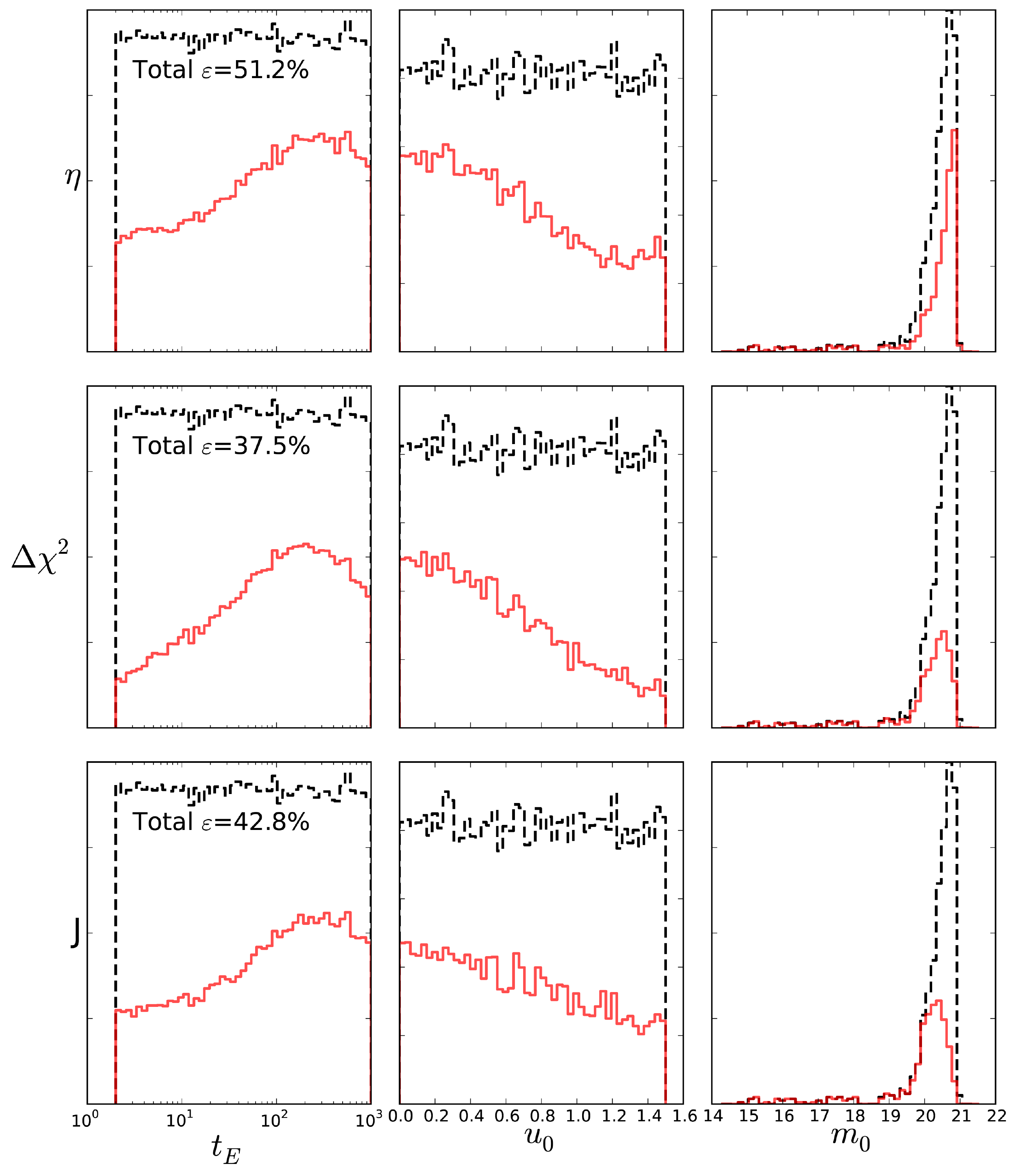}
	   \label{fig:subfig2}
	 }
\caption{Same as Figure~\ref{fig:detection_efficiency_4327}, for PTF field 4588.}\label{fig:detection_efficiency_4588}
\end{figure}

\begin{figure}
\centering
        \subfigure{
	   \includegraphics[width=.45\textwidth, trim=0 8 6 5, clip] {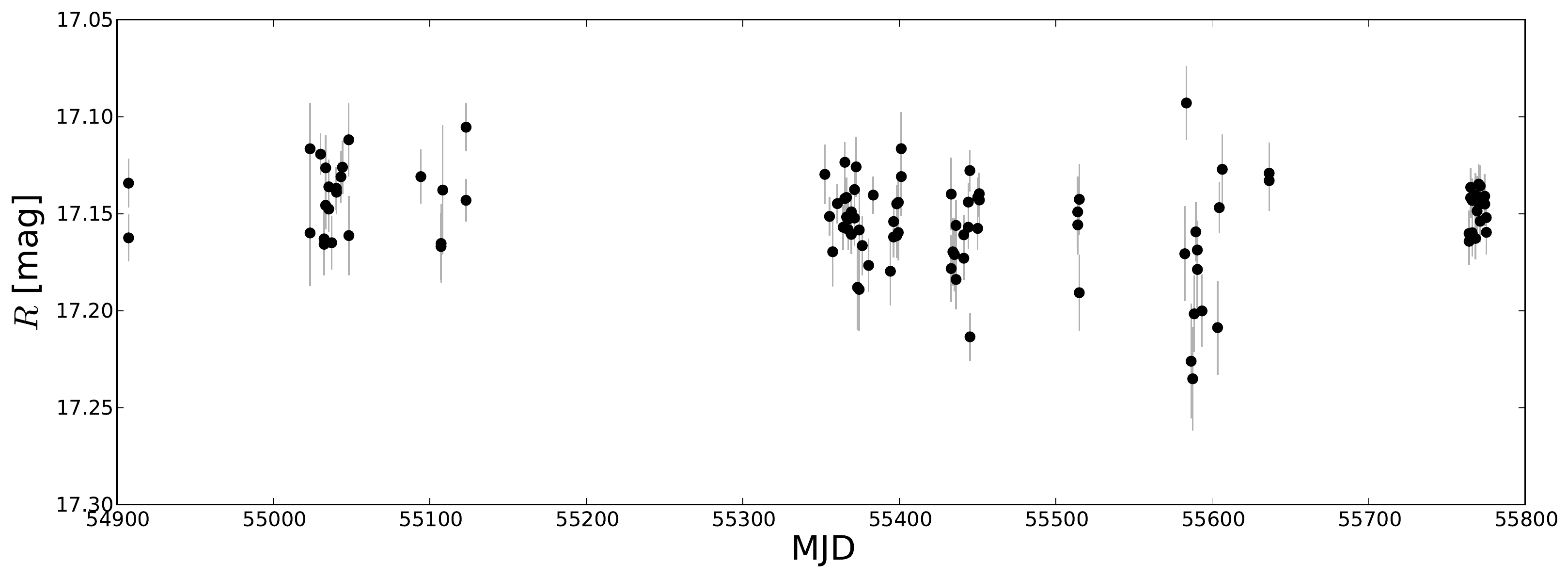}
	   \label{fig:subfig1}
	 }
	
	 \subfigure{
	   \includegraphics[width=.46\textwidth, trim=8 8 5 5, clip] {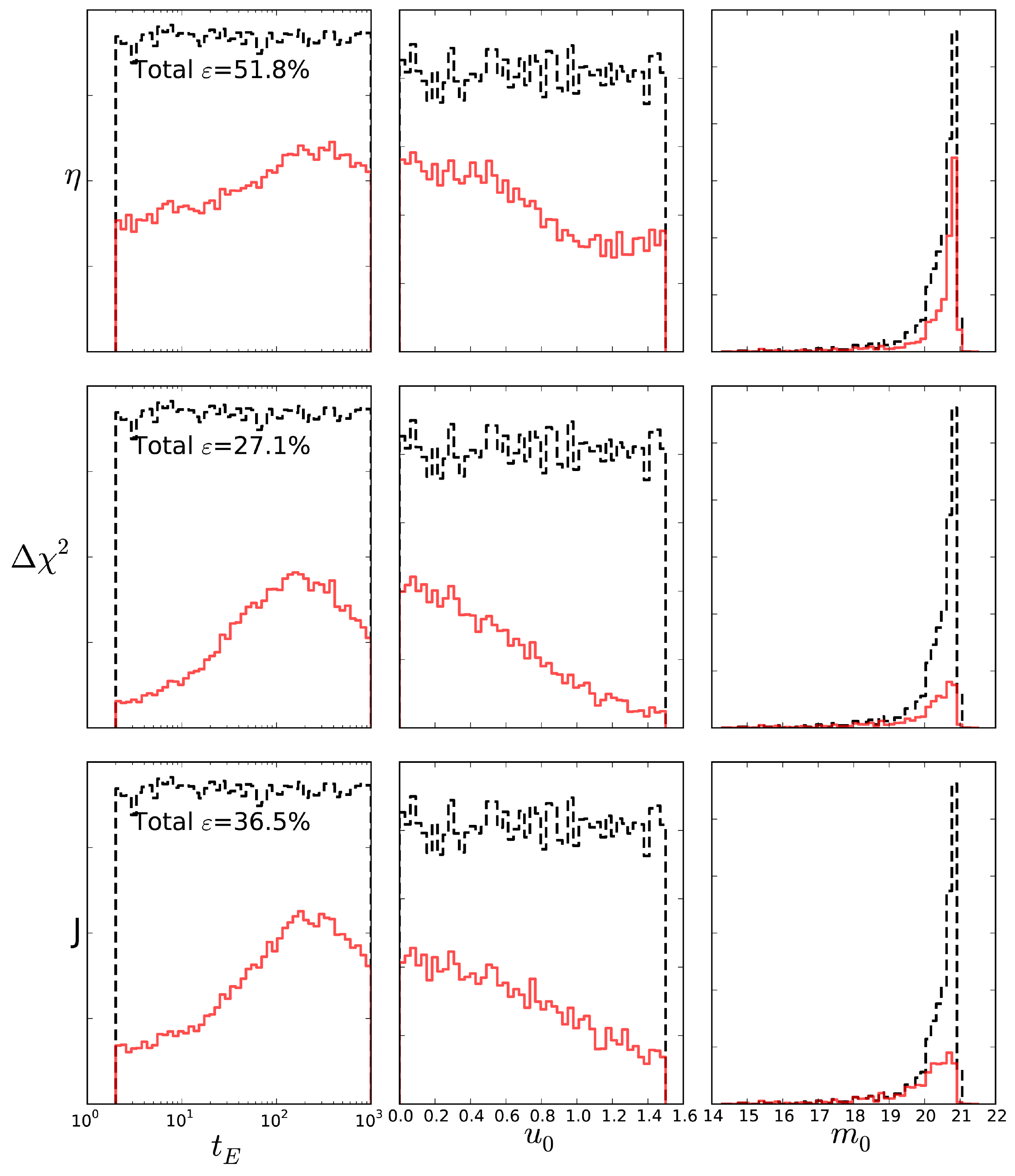}
	   \label{fig:subfig2}
	 }
\caption{Same as Figure~\ref{fig:detection_efficiency_4327}, for PTF field 100152.}\label{fig:detection_efficiency_100152}
\end{figure}

Figures~\ref{fig:detection_efficiency_4327}-\ref{fig:detection_efficiency_100152} show the results for three representative fields with different sampling patterns (top panels) and detection efficiency curves computed using this simulation (bottom panels). We ignore the indices \textit{K} and $\sigma/\mu$ because their integrated detection efficiencies are below 1\%. At a fixed FPR of 1\%, $Con$ performs poorly, but it may be useful as an initial cut if a higher FPR is used. In this work, we only consider a single-index selection method and thus reject $Con$. 

For each of these fields we find that $\eta$ consistently performs better than the other indices at recovering microlensing events. This is especially interesting because computing $\eta$ is $\sim$100$\times$ faster per light curve than computing $\Delta\chi^2$. Identifying candidate microlensing events in the full PTF dataset using this statistic is therefore computationally plausible.
	


\section{Searching for events in the full PTF dataset}\label{sec:search}
To identify candidate microlensing events we: 
\begin{enumerate}
	\item select the light curves from one of the 2097 fields with $>$10 $R$-band observations;
	\item identify those light curves with more than 50\% good data points and $>$10 good observations and run Monte Carlo simulations to compute the 1\% FPR limiting values for $\eta$, $\Delta \chi^2$, and $J$ for that field;
	\item compute $\eta$ for all the light curves in the field. While $\sim$10\% of the light curves pass the quality cut described above, only $\sim$0.1-0.5\% of these survive the FPR cut determined from the Monte Carlo simulations;
	\item for fields with SDSS coverage, we use Data Release 9 photometry \citep{dr9paper} to remove sources typed as Galaxy or QSO (using \texttt{objc\_type}\footnote{\url{http://www.sdss3.org/dr9/algorithms/classify.php}}) and flag candidate quasars \citep[using the cuts described in][]{richards02}; this eliminates $\sim$90\% of the remaining light curves;
	\item fit a microlensing event model and subtract the model, then recompute $\eta$ for all the selected light curves;
	\item reject the light curves for which the new value of $\eta$ still passes the cut\footnote{These are probably periodic variables.} --- this step eliminates $\sim$90-95\% of the remaining light curves;
	\item retain any surviving light curves for further inspection. Typically, there is $\lapprox$1 such light curve per field.
\end{enumerate}

This procedure identifies 2377 candidates from among the initial sample of $1.1\times10^9$ light curves. We search SIMBAD\footnote{This research has made use of the SIMBAD database, operated at CDS, Strasbourg, France.} and remove any candidate among these with a known extragalactic counterpart within 10\asec. For those in the SDSS footprint, we examine the SDSS images \citep{sdss_images} to identify extended objects. Removing extragalactic objects in this manner from our candidate list reduces contamination by, e.g., supernovae. 

\begin{figure*}[htb]
\centering
\includegraphics[width=2.1\columnwidth]{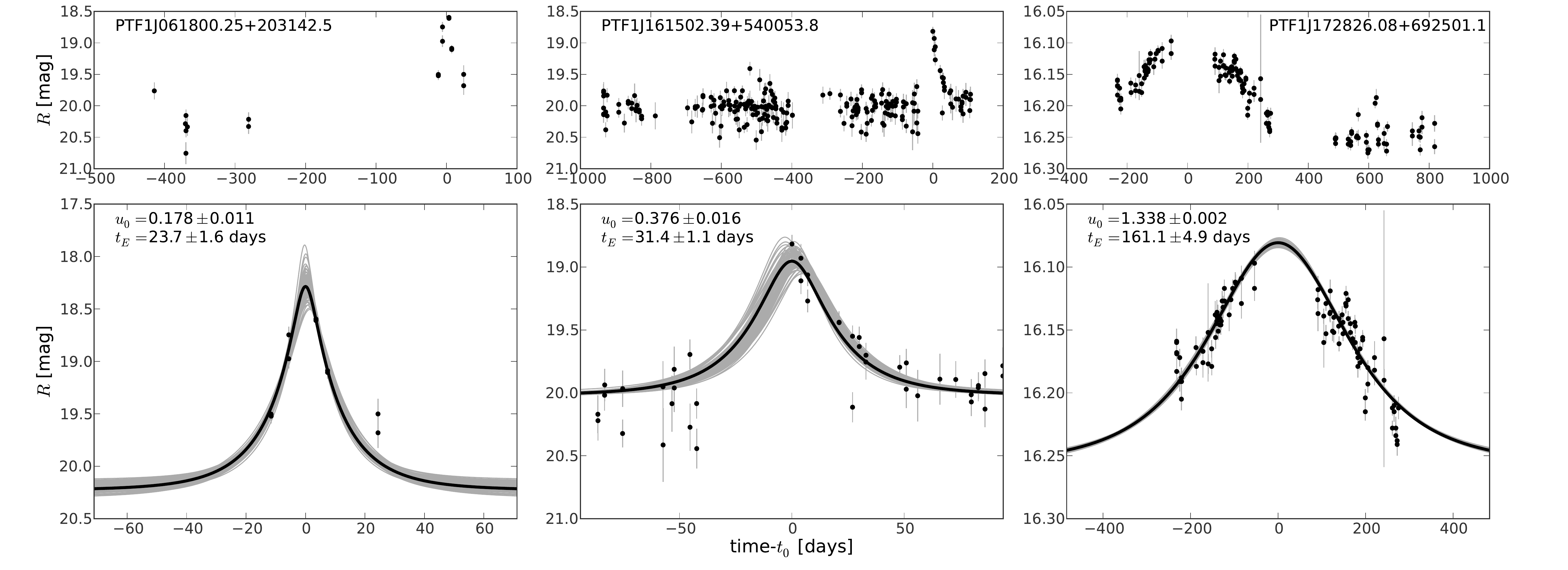}
\caption{Full light curves (top) and zooms around the transient maximum (bottom) for the three plausible microlensing event candidates. Bad (flagged) data points have been removed. Lines (gray, thin) show models with parameters sampled from the posterior probability distribution over the four parameters in the point-lens, point-source microlensing event model. Black (thick) line shows the maximum {\it a posteriori} model. }\label{fig:candidates}
\end{figure*}

\begin{deluxetable*}{cccccccccc}[!th]
\tablecaption{Photometric Properties of the Microlensing Candidates \label{table:photometry}}
\tablehead{
\colhead{PTF1} & \colhead{$R$}  & \colhead{$(g'-R)$}   & \colhead{$J$}   & \colhead{$(J-H)$} & \colhead{$(H-K)$} &  \colhead{$W1$}  & \colhead{$(W1-W2)$}  & \colhead{$(W2-W3)$} & \colhead{$(W3-W4)$}    
}
\startdata
J0335\tablenotemark{a} & 18.28$\pm$0.12 & \nodata & 16.59$\pm$0.13 & 0.73$\pm$0.18 & 0.79$\pm$0.18 & 15.06$\pm$0.04 & 0.18$\pm$0.09 & \nodata & \nodata \\
{\bf J0618} & 20.13$\pm$0.10 & 1.13$\pm$0.22 & 16.23$\pm$0.09 & 0.77$\pm$0.15 & 0.70$\pm$0.15 & 14.08$\pm$0.03 & 0.49$\pm$0.05 & 3.22$\pm$0.09 & 2.65$\pm$0.19 \\
J1206\tablenotemark{b} & 19.24$\pm$0.27 & \nodata & \nodata & \nodata & \nodata & \nodata & \nodata & \nodata & \nodata \\
J1315\tablenotemark{a} & 19.69$\pm$0.28 & \nodata & \nodata & \nodata & \nodata & 15.37$\pm$0.04 & 0.81$\pm$0.06 & 2.46$\pm$0.17 & 2.96$\pm$0.37 \\
J1532\tablenotemark{b} & 19.25$\pm$0.21 & \nodata & \nodata & \nodata & \nodata & 16.20$\pm$0.05 & $-0.43$$\pm$0.21 & \nodata & \nodata \\
{\bf J1615} & 20.01$\pm$0.24 & 1.10$\pm$0.29 & \nodata & \nodata & \nodata & 16.36$\pm$0.04 & 0.09$\pm$0.11 & \nodata & \nodata \\
J1716\tablenotemark{a} & 20.27$\pm$0.20 & 0.17$\pm$0.28 & \nodata & \nodata & \nodata & 15.56$\pm$0.04 & 0.85$\pm$0.07 & 2.93$\pm$0.15 & \nodata \\
{\bf J1728} & 16.20$\pm$0.10 & 0.82$\pm$0.10 & 14.70$\pm$0.04 & 0.58$\pm$0.06 & 0.16$\pm$0.09 & 13.92$\pm$0.02 & 0.01$\pm$0.03 & $-$0.28$\pm$0.50 & \nodata \\
J1733\tablenotemark{b} & 20.15$\pm$0.25 & 0.00$\pm$0.34 & 17.18$\pm$0.21 & \nodata & \nodata & \nodata & \nodata & \nodata & \nodata \\
J1747\tablenotemark{a} & 17.37$\pm$0.10 & 0.20$\pm$0.12 & 16.12$\pm$0.07 & 0.49$\pm$0.13 & 0.92$\pm$0.13 & 13.10$\pm$0.03 & 1.01$\pm$0.03 & 2.96$\pm$0.04 & 2.68$\pm$0.06 \\
J1933\tablenotemark{a} & 17.53$\pm$0.10 & 1.13$\pm$0.27 & 15.77$\pm$0.07 & 0.74$\pm$0.11 & 0.59$\pm$0.13 & 14.16$\pm$0.03 & 0.15$\pm$0.04 & 0.94$\pm$0.44 & \nodata
\enddata
\tablenotetext{a}{Photometry and/or visual inspection of the PTF/WISE images rule these out as plausible candidate events.}
\tablenotetext{b}{Available data are insufficient to determine the likelihood that these are plausible candidate events.}
\tablecomments{Objects in bold are plausible microlensing candidates.}
\end{deluxetable*}

We visually inspect each of the remaining $\sim$2000 light curves. We identify an additional $\sim$1100 objects with unknown long-term variability (e.g, Mira-type variables, quasars not in the SDSS footprint). An additional $\sim$600 light curves have bad or poorly calibrated data that mimics a transient increase in flux: diffraction spikes, ghosts, and scattered light can cause non-Gaussian, seeing-dependent variability. Where appropriate, as an additional test, we examine the PTF images to verify the data quality. A small fraction of the light curves are variable stars that survived our periodic variability cut. 

We are left with $\sim$300 unclassified transients. Through visual inspection of these light curves, we classify the bulk of these as novae, supernovae, flares, or outbursting systems, and identify 11 microlensing event candidates. We then search the literature for additional photometric data for these 11 objects. Most have a counterpart in WISE \citep{WISE}, while a few have a counterpart in 2MASS \citep{cutri03}; see Table~\ref{table:photometry} for a summary. We also obtained spectroscopy for several of these candidates. Below we detail what can be learned from these data. Generally, additional deeper optical/infrared imaging and/or optical spectroscopy is required to draw firm conclusions about the nature of the object.

\begin{figure}
\centering
\includegraphics[width=1.0\columnwidth] {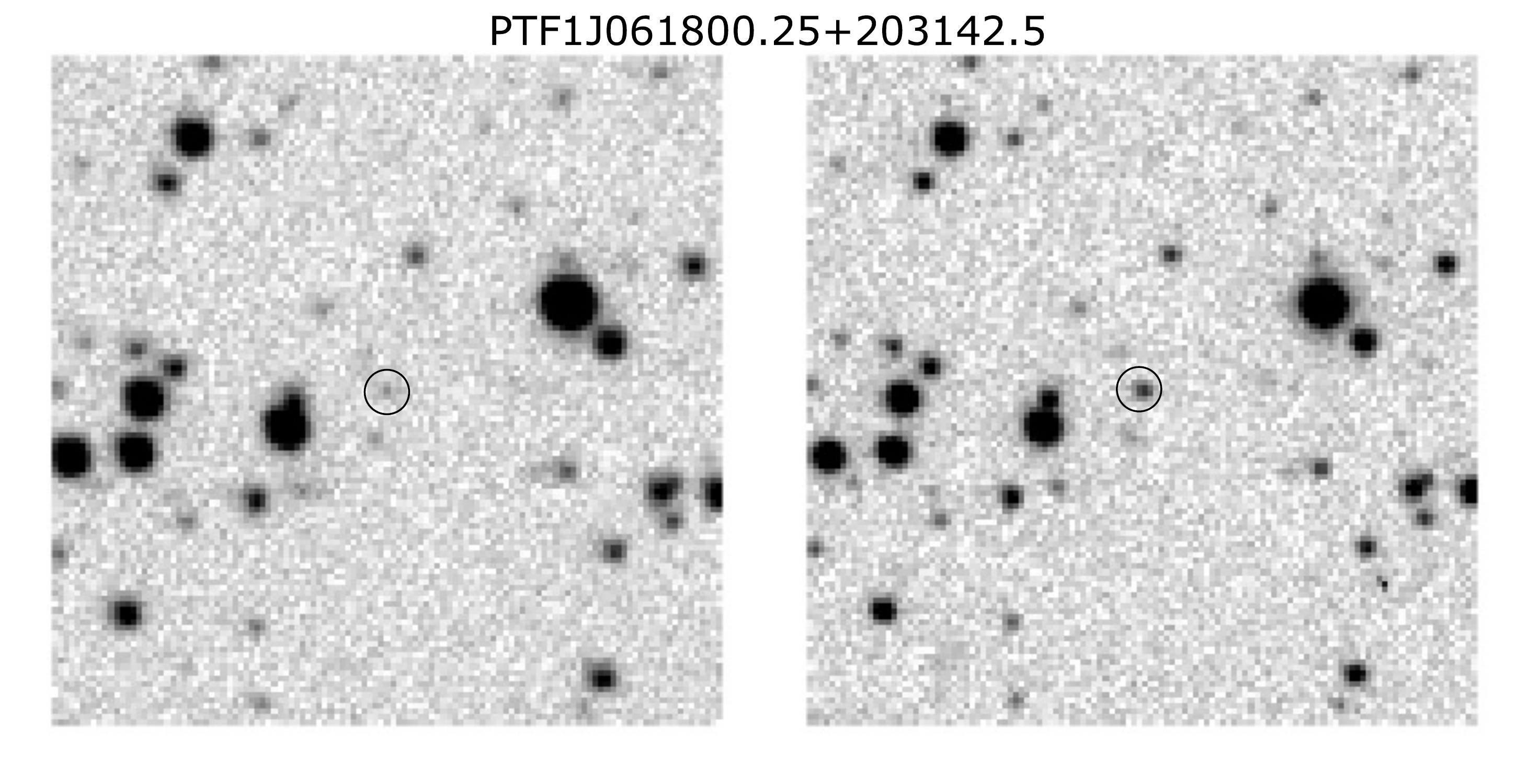}
\includegraphics[width=1.0\columnwidth] {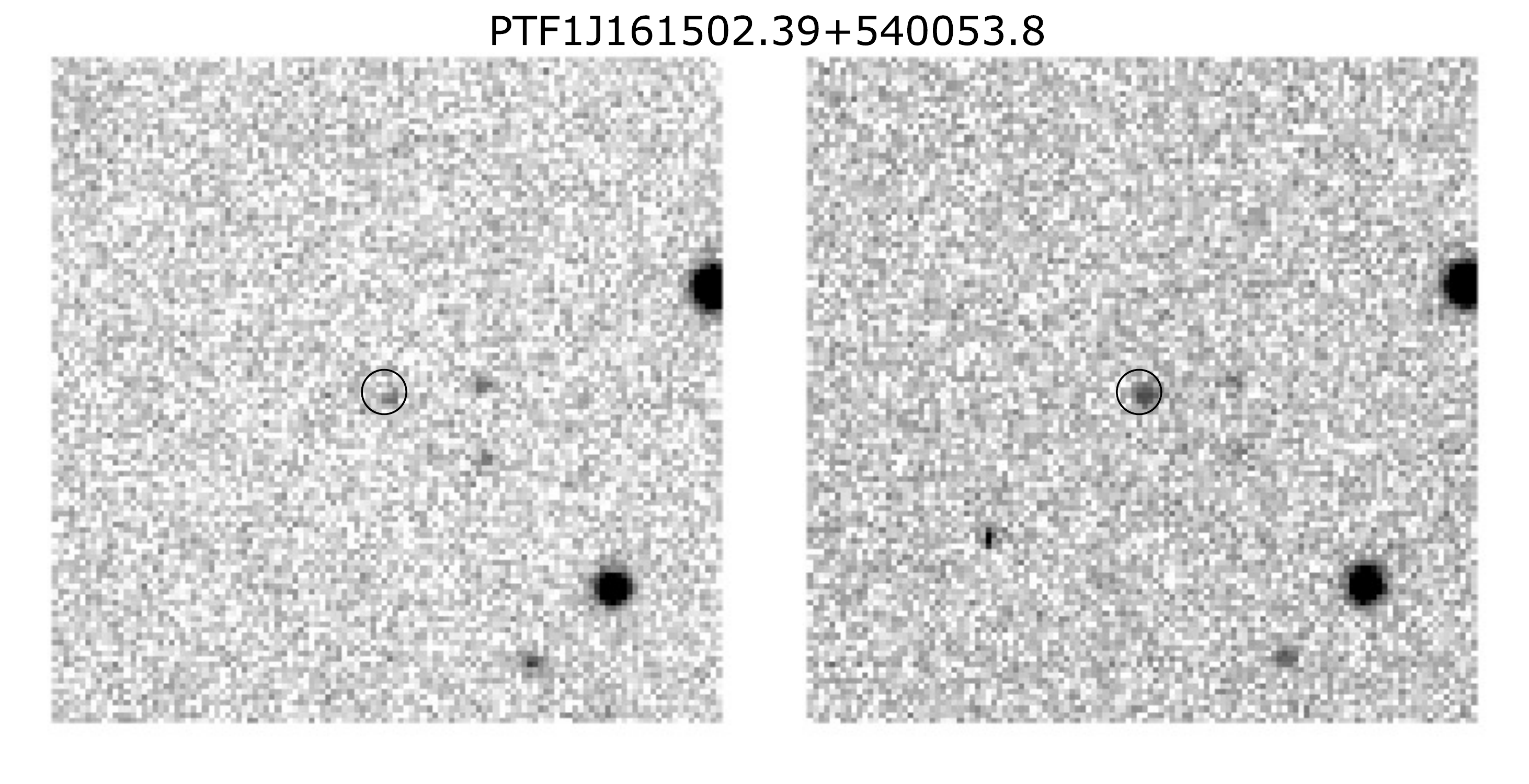}
\includegraphics[width=1.0\columnwidth] {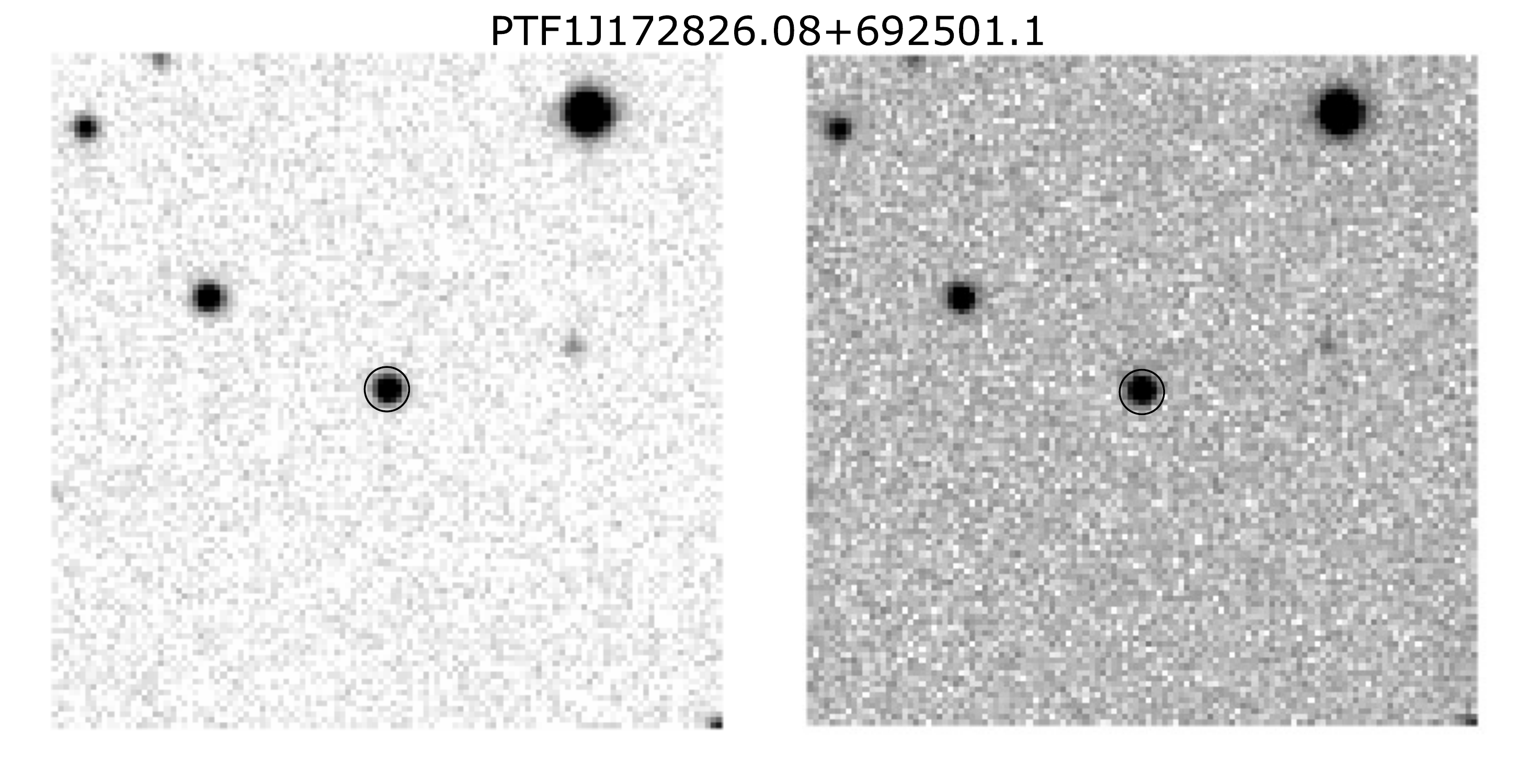}
\caption{$2'$ by $2'$ PTF $R$-band images of the three sources in Figure~\ref{fig:candidates}. {\it Left} ---  Sources at quiescence.  {\it Right} --- PTF image closest in time to the peak event time $t_0$ obtained by our model fits. Images are shown North-up, East-left.}\label{fig:postage_stamps}
\end{figure}

To summarize what follows: we rule out five of the 11 candidates as being likely extragalactic transients based on photometry and/or imaging from WISE, 2MASS, and/or PTF. Three of the candidates are not detected in WISE and/or 2MASS; follow-up imaging or spectroscopy is needed to draw any conclusions about these objects. The remaining three objects (PTF1J061800.25$+$203142.5, J161502.39$+$540053.8, and J172826.08$+$692501.1) are likely stars and, given the observed PTF variability, plausible microlensing events. We emphasize, however, that lacking simultaneous, multi-color photometry during the events, we cannot claim these as microlensing events with any confidence. These three candidates are shown in Figure~\ref{fig:candidates}, and PTF cutouts of the associated sources at quiescence and near peak brightness are shown in Figure~\ref{fig:postage_stamps} (light curves for the other eight events are shown in Figure~\ref{fig:not_candidates}).

For these three plausible events, we use a Markov-chain Monte-Carlo algorithm \citep{goodman, dfm} to derive posterior probability distributions over each parameter in the point-source, point-lens microlensing model (Eq.~\ref{eq:ml_model}). Overplotted on Figure~\ref{fig:candidates} are samples from these posterior distributions along with the maximum {\it a posteriori} (MAP) model; the corresponding MAP event parameters are listed in Table~\ref{table:candidates1}. The event durations for PTF1J0618 and PTF1J1615 are reasonable, given that a ${\sim}0.1\ M_\odot$ lens at ${\sim}500~\mathrm{pc}$ in the thick disk ($v_{tan}\sim 50~\mathrm{km}~\mathrm{s}^{-1}$) has a typical event duration $t_E \sim 20~\mathrm{days}$, but the significantly longer duration of PTF1J1728 ($>$150 days) could be a sign of a greater mass lens or larger distance.

PTF1J0618 and PTF1J1728 have counterparts in 2MASS, and we can therefore use the \citet{kev07} $J$-band absolute magnitudes for stars of the relevant spectral types to estimate rough distances to these stars, assuming they have solar metallicity (see below for discussion of how we obtain spectral types for these stars). For PTF1J0618, this results in a distance that ranges from 1.1 kpc (if it is a M1 star) to 0.8 kpc (if it is a M3 star), corresponding to heights above the Galactic Plane of 43 and 31 pc, respectively. For PTF1J1728, assuming it is a K0-K5 star results in distances ranging from 1.3 to 0.7 kpc, and heights of 0.9 to 0.5~kpc. 

We lack a spectrum for PTF1J1615, so the distance estimate to this star is even more uncertain. Still, we compare the PTF $R$ magnitude to the stellar library compiled by \citet{Pickles1998} to estimate a distance, assuming (from its photometry) that it is a K5-M2 star.\footnote{This ignores differences between Mould and Cousins $R$ filters.} This results in distances ranging from 4.2 to 1.9 kpc and heights from 2.3 to 1.0~kpc. The estimated heights above the Galactic Plane suggest that PTF1J0618 is a thin disk star, while the other two, PTF1J1615 and PT1J1728, are plausible thick disk members \citep{Bochanski2010}.

\begin{deluxetable*}{ccccc}[!h]
\tabletypesize{\scriptsize}
\tablecaption{Derived Microlensing Event Parameters for Plausible Candidates \label{table:candidates1}}
\tablehead{
\colhead{PTF1} & \colhead{$u_0$}  & \colhead{$t_E$ [days]}   & \colhead{$t_0$ [MJD]}   & \colhead{$m_0$ [R]}     
}
\startdata
J0618 & 0.178$\pm$0.011 & 23.7$\pm$1.6 & 55880.0$\pm$0.4 & 20.222$\pm$0.034 \\
J1615 & 0.376$\pm$0.016 & 31.4$\pm$1.1 & 55895.5$\pm$2.1 & 20.019$\pm$0.004 \\
J1728 & 1.338$\pm$0.002 & 161.1$\pm$4.9 & 55192.2$\pm$1.4 & 16.260$\pm$0.002 
\enddata
\end{deluxetable*}

\subsection{Plausible Microlensing Candidates}
\subsubsection*{PTF1J061800.25$+$203142.5} 
\begin{figure}[!h]
	\centering
	\includegraphics[width=.48\textwidth, trim=0 0 15 40, clip] {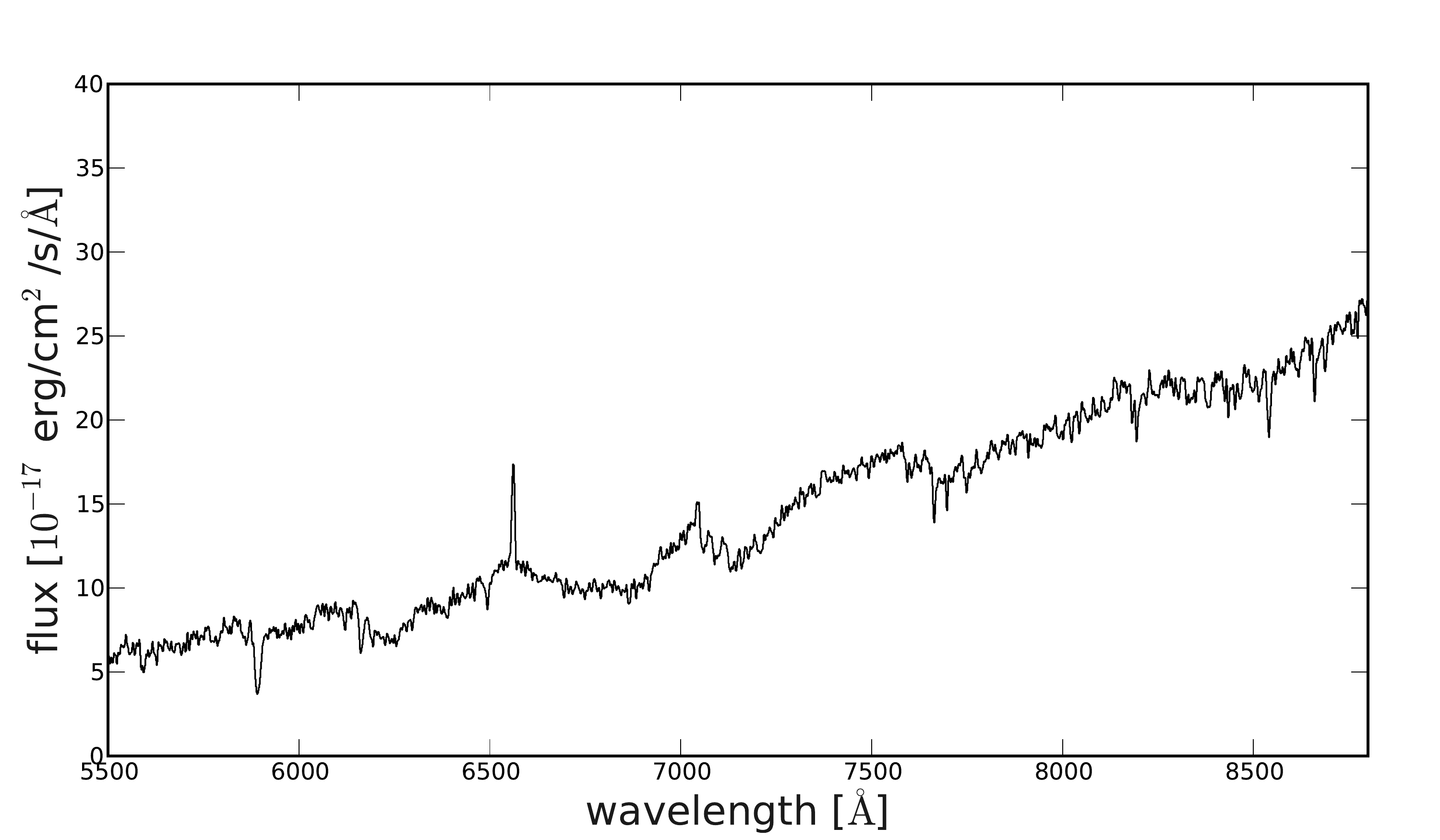}
\caption{\label{fig:1206i_spectrum} DBSP spectrum of PTF1J0618, an early M dwarf star, smoothed with a Gaussian filter ($\sigma = 1$). The timescale of the event detected by PTF is inconsistent with those generally observed in flares, which are the mostly likely contaminants in identifying microlensing events involving M stars.}
\end{figure}

PTF1J0618 is detected in 2MASS and in all four WISE filters. The corresponding colors are difficult to interpret, however: the object's $(J-H)$ color is consistent with that of a mid-M dwarf \citep{kev07}, but its $(H-K)$ is significantly redder than expected for such a star. Meanwhile, PTF1J0618's WISE colors suggest it is an extragalactic object \citep[cf.\ Figure 14 in][]{yan2013}. We observed PTF1J0618 for 1200~s with the Double Spectrograph \citep[DBSP;][]{oke_gunn} on the Hale 5-m telescope at Palomar Observatory, CA, on 2013 Feb 19; see Figure~\ref{fig:1206i_spectrum}. The spectrum was obtained with the D55 dichroic; from the atmospheric cutoff to 5500~\AA, the grating had 600 line~mm$^{-1}$ and was blazed at 4000~\AA, giving a resolution of 1.1~\AA. From 6300-8800~\AA, the grating was 158 line~mm$^{-1}$, was blazed at 7500~\AA, and gave a resolution of 2.5~\AA. The data were reduced using standard IRAF routines.\footnote{IRAF is distributed by the National Optical Astronomy Observatories, which are operated by the Association of Universities for Research in Astronomy, Inc., under cooperative agreement with the National Science Foundation.}

We analyzed this spectrum with the HAMMER IDL package \citep{kev07}. The HAMMER measures a suite of spectral features and provides automated spectral types by comparison to templates. The typical uncertainty in HAMMER spectral typing is $\lesssim$1 spectral subclass. PTF1J0618 is clearly an early M star, with a spectral type of M2-M3. 

M stars are well-known sources of stellar flares, and in sparse data these could easily be confused for a microlensing event. However, typical flares on M stars last minutes to hours, with the longest seen being of order eight hours \citep{kowal2010}. By contrast, the candidate event detected by PTF had a duration $\sim$30 days and is therefore highly unlikely to have been a flare.

Could PTF1J0618 be an outbursting symbiotic star? To test this hypothesis, we compare the DBSP spectrum to a set of M1-M2 dwarf and giant star spectra. We inspect the depth of the Ca II triplet lines and also calculate several of the spectral indices collected by \citet{kev06} to distinguish between dwarfs and giants (e.g., the fluxes in the Na D 5900 and CN 7900 bands). These all indicate that PTF1J0618 is a dwarf star, rendering it unlikely that the PTF event was a symbiotic-type outburst. We further note that those outbursts typically last $\sim$years \citep[e.g.,][]{tomov2013}. We conclude that this is a plausible candidate microlensing event.


\subsubsection*{PTF1J161502.39$+$540053.8} 
PTF1J1615 is one of two candidates that fall in the SDSS footprint; its counterpart, SDSS J161502.43$+$540053.6, has a proper motion of 2.66 mas/year, consistent with its automated SDSS classification as a star. The star has $u = 22.9$ mag,\footnote{These are SDSS PSF magnitudes. Typically, PSF fitting provides better estimates of isolated star magnitudes; see \citet{stoughton02}.} beyond the $u = 22.0$ 95\% completeness limit for SDSS \citep{stoughton02}, and we therefore focus on its $griz$ photometry. With $(g-r) = 1.10\pm0.05$, $(r-i) = 1.00\pm0.03$, and $(i-z) = 0.56\pm0.04$, the star's colors are consistent with those of a late K/early M star \citep{kev07}. PTF1J1615 has no counterpart in 2MASS, and is only detected in the WISE $W1$ and $W2$ bands, making it hard to draw any conclusions based on these data. Still, its $(W1-W2)$ is consistent with what is seen for stars \citep[cf.\ Figure 14 in][]{yan2013}. Here again, the most likely source of contamination is flaring. However, the event detected by PTF had a duration $\sim$20 days, and so is highly unlikely to have been a flare. We conclude that this is a plausible candidate microlensing event.

\subsubsection*{PTF1J172826.08$+$692501.1} 
PTF1J1728 is the other candidate in the SDSS footprint. Its SDSS match, SDSS J172826.07$+$692501.3, has a proper motion of 1.81 mas/year, consistent with its automated classification as a star. PTF1J1728 also has a counterpart in 2MASS, and its SDSS/2MASS colors --- $(u-g) = 1.81\pm0.04$, $(g-r) = 0.79\pm 0.03$, $(r-i) = 0.35\pm0.03$, $(i-z) = 0.21\pm0.03$, $(z-J) = 0.97 \pm0.05$, $(J-H) = 0.58\pm0.06$, and $(H-K) = 0.16\pm 0.09$ --- are consistent with those of a K star \citep{kev07}. Its $(W1-W2)$ and $(W2-W3)$ colors place PTF1J1728 in the stellar locus in Figure 14 of \citet[][]{yan2013}. 

\begin{figure}[!h]
	\centering
	\includegraphics[width=.48\textwidth, trim=0 0 15 40, clip] {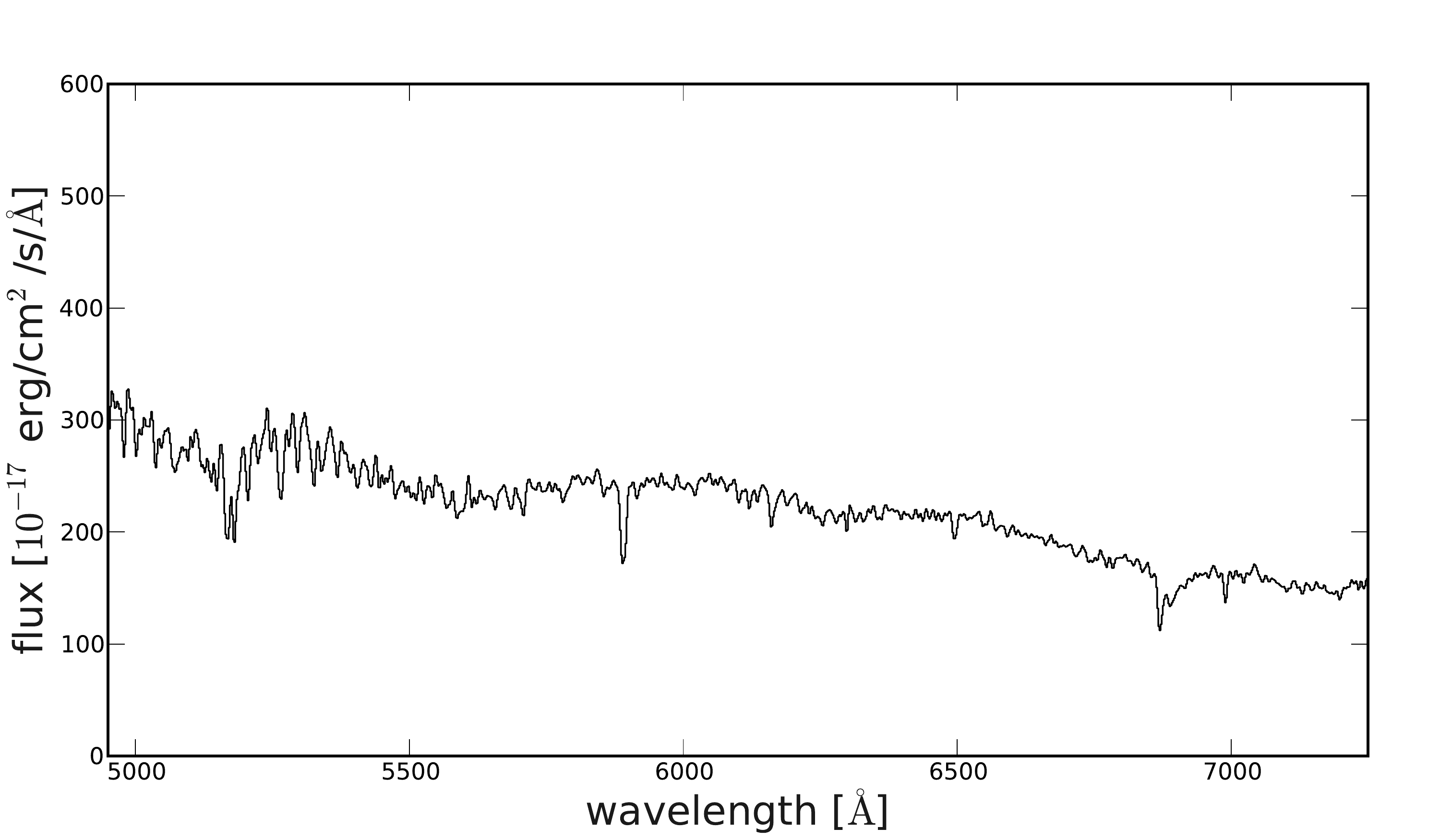}
\caption{\label{fig:1217cv_spectrum} MODspec spectrum of PTF1J172826.08$+$692501.1, a mid-K star.}
\end{figure}

E.~Bowsher and S.~Douglas obtained two 1200-s spectra of PTF1J1728 with the MDM Observatory Modular Spectrograph (MODspec) on the Hiltner 2.4-m telescope at MDM Observatory, Kitt Peak, AZ, on 2012 Nov 14.\footnote{The MDM Observatory is operated by Dartmouth College, Columbia University, Ohio State University, Ohio University, and the University of Michigan.} MODspec was configured to provide coverage from 4500 to 7500 \AA~with $\sim$1.8 \AA~sampling and a spectral resolution of $\sim$3300. The data reduction was performed by S.~Douglas using standard IRAF routines. 
All the spectra from this night suffered from significant noise blueward of 5000 \AA~and redward of 7000 \AA; in  Figure~\ref{fig:1217cv_spectrum}, we therefore present the spectrum over that 2000 \AA~range. While the spectrum is noisy, PTF1J1728 appears to be a mid-K star; its HAMMER spectral type is K3. We conclude that this is a plausible candidate microlensing event.

\subsection{Candidates Ruled out by Photometry and/or Imaging}
\subsubsection*{PTF1J033545.80$-$041849.1} 
PTF1J0335 has a counterpart in 2MASS, and its $(J-H)$ and $(H-K)$ colors indicate that it is unlikely to be a star \citep{kev07}. Because it is only detected in the $W1$ and $W2$ bands, the WISE color does not help much: it is consistent with that of stars and of compact galaxies \citep[cf.\ Figure 14 in][]{yan2013}. We tentatively conclude that this not a plausible microlensing candidate. 

\subsubsection*{PTF1J131500.09$+$715032.5} 
PTF1J1315 has no counterpart within 10\asec\ in 2MASS, but is detected in all four WISE bands. 
The $W3$ image appears to have an offset nucleus compared to other filters; this could be a background source, and we therefore do not use the $W3$ magnitude because of this possible contamination. The $(W1-W2)$ color places the object well away from the stellar locus \citep[cf.\ Figure 14 in][]{yan2013}. We tentatively conclude that this is a likely extragalactic source and therefore not a plausible microlensing candidate. 

\subsubsection*{PTF1J171606.82$+$474423.7} 
PTF1J1716 does not have a counterpart in 2MASS. Its $(W1-W2)$ and $(W2-W3)$ colors are consistent with this object being a QSO \citep[cf.\ Figure 14 in][]{yan2013}. We tentatively conclude that this not a plausible microlensing candidate.
 
\subsubsection*{PTF1J174736.51$+$300506.9} 
PTF1J1747 has counterparts in both 2MASS and WISE. The 2MASS colors suggest that this is not a star \citep{kev07}, while the WISE colors are consistent with those of a QSO \citep[cf.\ Figure 14 in][]{yan2013}. 
We tentatively conclude that this not a plausible microlensing candidate. 


\subsubsection*{PTF1J193330.11$+$451501.1} 
PTF1J1933 is detected in both 2MASS and WISE. The 2MASS colors suggest that this is not a star \citep{kev07}, while the WISE colors are consistent with PTF1J1933 being an elliptical galaxy \citep[cf.\ Figure 14 in][]{yan2013}. This interpretation is strengthened by a visual inspection of the PTF and WISE images, which show that the object is elongated in the Northeast-Southwest direction. We tentatively conclude that this not a plausible microlensing candidate. 

\subsection{Candidates Lacking Sufficient Data to Evaluate Likelihood}
\subsubsection*{PTF1J120642.60$-$192016.4} 
PTF1J1206 has no counterpart within 10\asec\ in 2MASS. While a counterpart is listed in the WISE data 3\asec\ from the PTF position, a visual comparison of the co-added PTF $R$-band images and of the WISE images indicates that the PTF object may be blended with other objects in the WISE data or not detected at all. However, the WISE photometric flags for this nearby source suggest it is clean and unblended. The available information is insufficient to draw even tentative conclusions about the nature of this object.

\subsubsection*{PTF1J153202.91$+$674825.1} 
PTF1J1532 has no counterpart in 2MASS, but is detected in the WISE $W1$ and $W2$ bands. The corresponding $(W1-W2) = -0.43\pm0.2$ color is somewhat unusual, as most objects appear to have $(W1-W2)\ \gapprox\ 0.0$, with compact galaxies being the rare exceptions \citep[cf.\ Figure 14 in][]{yan2013}. Here again, we cannot draw any conclusions about the nature of this object. 

\subsubsection*{PTF1J173301.07$+$374311.9} 
PTF1J1733 is detected in the 2MASS $J$ and $K$ bands (but not $H$). There is no object in the WISE catalog within 10\asec\ of PTF1J1733. Although the $(J-K$) color appears inconsistent with PTF1J1733 being a star \citep{kev07}, we cannot draw any firm conclusions about the nature of this object. 

\section{Conclusions and Future Work}\label{sec:conc}
Using time-domain data from the Palomar Transient Factory, we have developed a new method for identifying interesting transients in massive light-curve data sets and for searching for sparse-field microlensing events. We have examined the detection efficiency of recovering simulated microlensing events using a set of variability statistics adapted from \citet{shin2009}. We determined selection criteria for each statistic using Monte Carlo simulations to simulate flat light curves with Gaussian noise, and by choosing the selection boundaries such that our cuts achieved a 1\% false positive recovery rate. We then simulated microlensing events in real PTF data and computed the detection efficiency for each statistic. 

We found that the von Neumann ratio, $\eta$ \citep[or Durbin-Watson statistic;][]{von_neumann1941, durbin50}, performs better than previously used statistics in recovering injected microlensing events in non-uniformly sampled data. We used $\eta$ to develop a selection procedure for extracting microlensing event candidates from the PTF light-curve archive. Among the $1.1 \times 10^9$ light curves with $>$10 $R$-band observations, we first identified $\sim$2000 interesting candidate transients, from which we selected 11 candidate microlensing events.

A large fraction of the contaminants among the $\sim$2000 interesting candidates were AGN or quasars exhibiting long-term, peaked variability. We also recovered a large number of transient events such as novae, outbursts, and flares. We have cross-referenced this list of objects with the PTF transient detection system and found that most of these objects are known sources, but we discovered at least two supernovae that the PTF pipeline had missed. 

Of the 11 candidate microlensing events, we tentatively ruled out eight by examining the available photometry and images, or because they lacked such data. We labeled the three remaining objects as plausible microlensing events, but lacking simultaneous multi-color imaging during the events, we cannot confirm these candidates as true microlensing events. 

\begin{figure*}
\centering
\includegraphics[width=2.1\columnwidth] {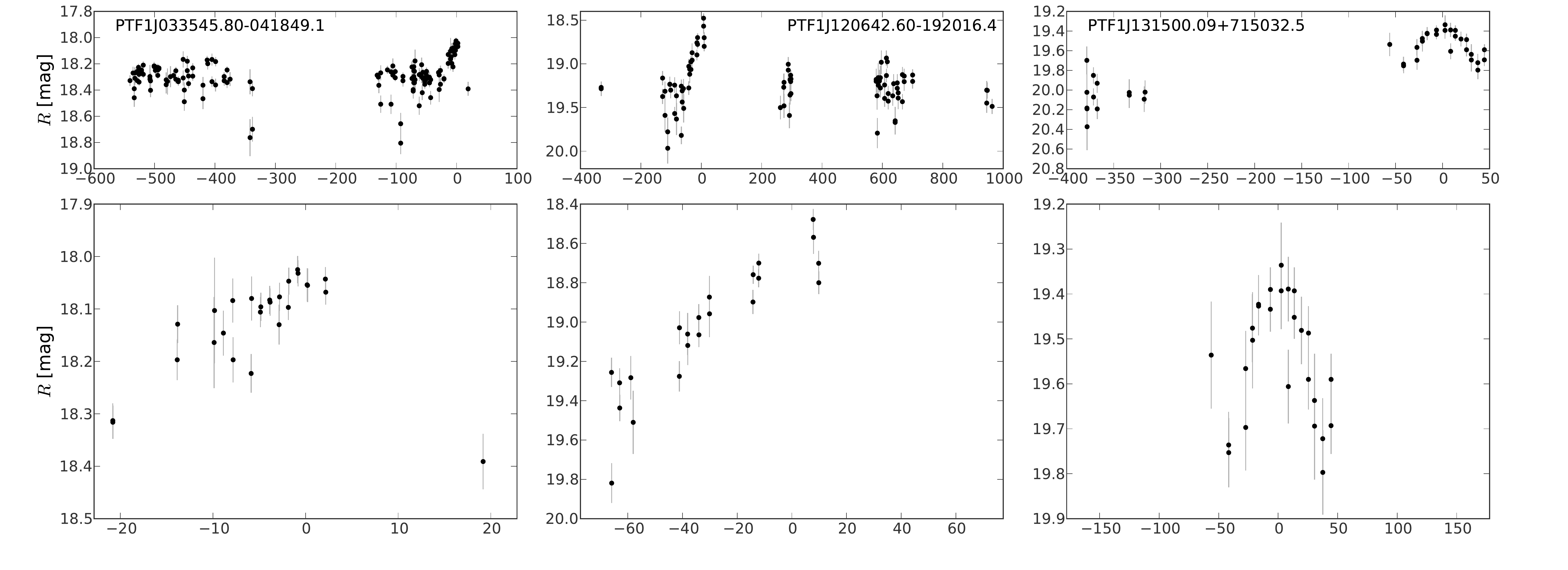}
\includegraphics[width=2.1\columnwidth] {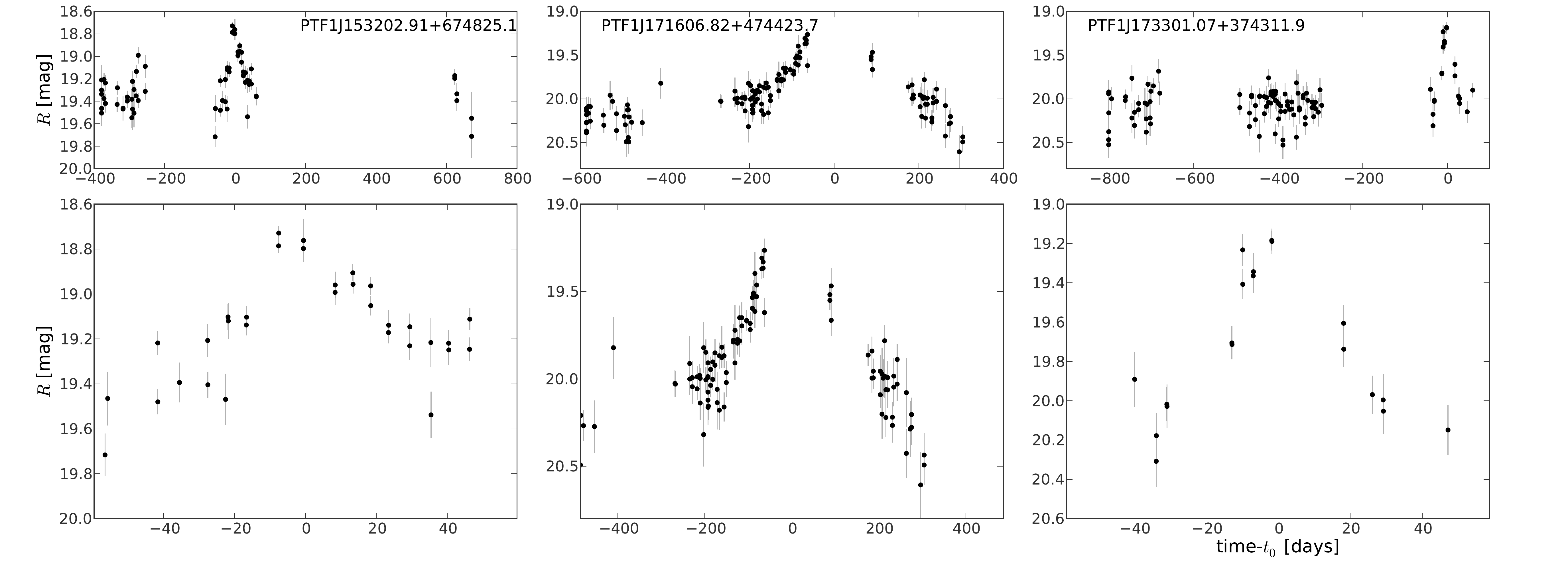}
\includegraphics[width=2.1\columnwidth] {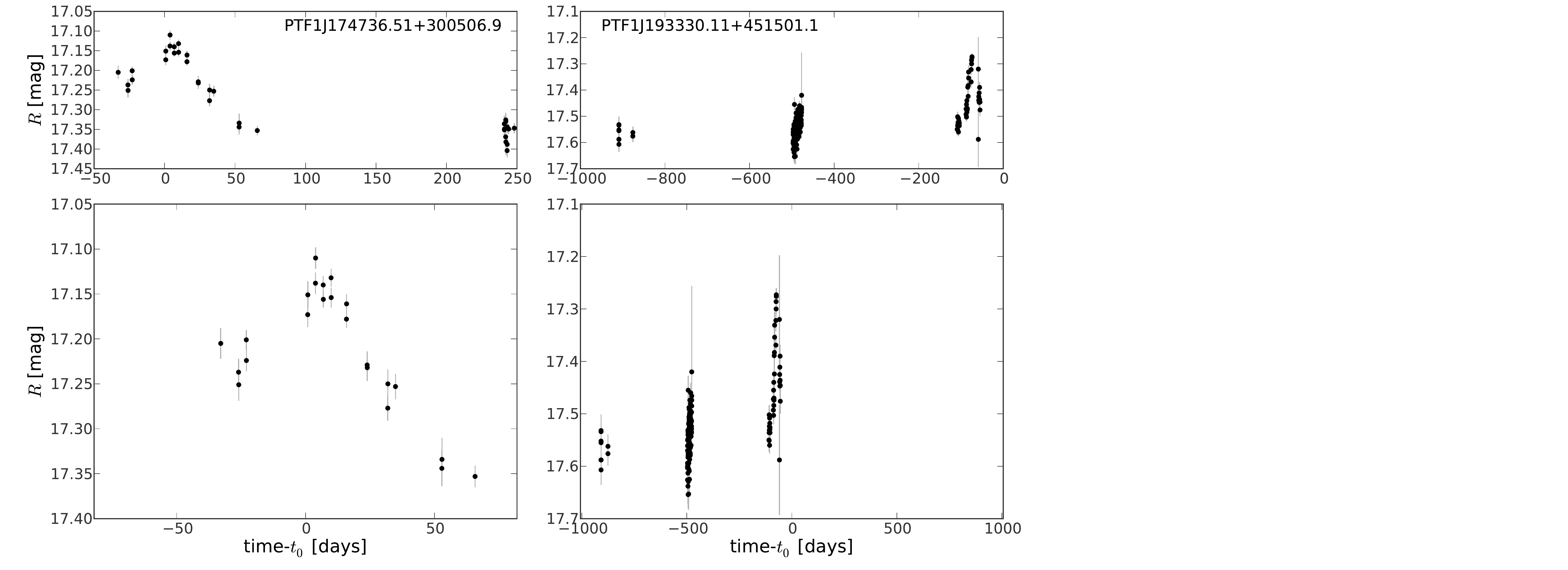}
\caption{Full light curves (top) and zooms around the transient maximum (bottom) for the rejected microlensing event candidates. Flagged data points have been removed.}\label{fig:not_candidates}
\end{figure*}

For these three plausible events, we use a Markov-chain Monte-Carlo algorithm to derive posterior probability distributions over each parameter in the point-source, point-lens microlensing model. The event durations for two of the candidates are reasonable, but the significantly longer duration of the third candidate could be a sign of a greater mass lens or larger distance. 

While this number of candidate events is consistent with simple predictions, microlensing event rate predictions away from the Galactic Plane are generally not well constrained. In a forthcoming paper, we will present a detailed investigation of the all-sky microlensing event rate,  generated from a model that includes realistic mass and velocity distributions
(Fournier et al., in prep). 

Our algorithm identified $\sim$2000 transient candidates from a database of over $10^9$ light curves, each of which we then visually inspected. Rigorously classifying and cataloging these sources was outside of the scope of this paper, but many  presented interesting forms of variability. Though it is not necessarily scalable for future surveys to look at all light curves selected by using our algorithm, these $\sim$2000 light curves, once carefully classified, could serve as a training set for a more sophisticated machine-learning approach \cite[e.g.,][]{bloomPasp, brink2013} to classify the light curves that survive our selection criteria. Approaches that combine statistical cuts such as those described in this paper and some type of machine learning will be essential to making the most of the future time-domain data such as that collected by the Large Synoptic Survey Telescope.

\acknowledgments
We thank E.~Bowsher, S.~Douglas, and S.~Tang for their assistance in obtaining spectra for our candidates, and M.~Modjaz for help in interpreting several of these spectra. We are grateful to the anonymous referee and to our editor, E.~Feigelson, for comments that improved the paper.

This paper is based upon work supported by a National Science Foundation Graduate Research Fellowship under Grant No.\ 11-44155 and on observations obtained with the Samuel Oschin Telescope as part of the Palomar Transient Factory project, a scientific collaboration between the California Institute of Technology, Columbia University, Las Cumbres Observatory, the Lawrence Berkeley National Laboratory, the National Energy Research Scientific Computing Center, the University of Oxford and the Weizmann Institute of Science. 

Funding for SDSS-III has been provided by the Alfred P. Sloan Foundation, the Participating Institutions, the National Science Foundation, and the U.S. Department of Energy Office of Science. The SDSS-III web site is \url{http://www.sdss3.org/}.

SDSS-III is managed by the Astrophysical Research Consortium for the Participating Institutions of the SDSS-III Collaboration including the University of Arizona, the Brazilian Participation Group, Brookhaven National Laboratory, University of Cambridge, Carnegie Mellon University, University of Florida, the French Participation Group, the German Participation Group, Harvard University, the Instituto de Astrofisica de Canarias, the Michigan State/Notre Dame/JINA Participation Group, Johns Hopkins University, Lawrence Berkeley National Laboratory, Max Planck Institute for Astrophysics, Max Planck Institute for Extraterrestrial Physics, New Mexico State University, New York University, Ohio State University, Pennsylvania State University, University of Portsmouth, Princeton University, the Spanish Participation Group, University of Tokyo, University of Utah, Vanderbilt University, University of Virginia, University of Washington, and Yale University. 

This publication makes use of data products from the Wide-field Infrared Survey Explorer, which is a joint project of the University of California, Los Angeles, and the Jet Propulsion Laboratory/California Institute of Technology, funded by the National Aeronautics and Space Administration.

This research made use of Astropy, a community-developed core Python package for astronomy \citep{astropy2013}.


\end{document}